\documentclass[12pt]{article}

\newcommand{\br}{{\bf {R}}}
\newcommand{\Om}{\Omega}

\newcommand{\tr}{\rm{Tr}}

\newcommand{\bs}{{\bf S}}

\usepackage{latexsym,amsfonts,amsmath,amsthm,amssymb}

\usepackage[T1]{fontenc}
\usepackage[latin1]{inputenc}
\title{Quantum mechanics as an asymptotic projection of statistical mechanics of classical fields:
derivation of Schr\"odinger's, Heisenberg's and von Neumann's
equations}
\author{Andrei Khrennikov\\
International Center for Mathematical
Modeling \\ in Physics and Cognitive Sciences,\\
University of V\"axj\"o, S-35195, Sweden}

\begin{document}
\maketitle

\begin{abstract}We show that QM can be represented as a natural
projection of a classical statistical model on the phase space
$\Omega= H\times H,$ where $H$ is the real Hilbert space.
Statistical states are given by Gaussian measures on $\Omega$ having
zero mean value and dispersion of very small magnitude $\alpha$
(which is considered as a small parameter of the model). Such
statistical states can be interpreted as fluctuations of the
background field, cf. with SED and Nelson's mechanics. Physical
variables (e.g., energy) are given by maps $f: \Omega \to {\bf R}$
(functions of classical fields). The conventional quantum
representation of our prequantum classical statistical model is
constructed on the basis of the Taylor expansion (up to the terms of
the second order at the vacuum field point $\psi_{\rm{vacuum}}\equiv
0)$ of variables $f: \Omega \to {\bf R}$ with respect to the small
parameter $\sqrt{\alpha}.$ The complex structure of QM is induced by
the symplectic structure on the infinite-dimensional phase space
$\Omega.$ A Gaussian measure (statistical state)
 is represented in QM by its covariation operator.
 Equations of Schr\"odinger, Heisenberg and von Neumann
are  images of Hamiltonian dynamics on $\Omega.$ The main
experimental prediction of our prequantum model is that experimental
statistical averages can deviate from ones given by QM.
\end{abstract}

\section{Introduction}
In the first part of this paper [1] we demonstrated that, in spite
of all ``NO-GO'' theorems, it is possible to construct a general
prequantum classical statistical model, cf. with SED [2], [3],
Nelson's stochastic mechanics [4] and Hooft's deterministic
prequantum models [5], [6]. The phase space of this model is the
infinite dimensional Hilbert space. Thus classical ``systems'' are
in fact classical fields. We call this approach {\it Prequantum
Classical Statistical Field Theory} (PCSFT). There was constructed a
natural map $T$ establishing the correspondence between classical
and quantum statistical models. This map $T$ produces the following
relation between classical and quantum averages:
\begin{equation}
\label{AVP} <f>_\rho = \alpha<T(f)>_{T(\rho)} + o( \alpha), \;
\alpha \to 0,
\end{equation}
where $\rho$ and $f$ are, respectively, a classical statistical
state and a classical variable. Here $ \alpha$ -- the dispersion of
the Gaussian measure $\rho$ (having zero mean value) --  is
considered as a small parameter of the model:
$$
\sigma^2(\rho) =\int \Vert \psi \Vert^2 d \rho (\psi)=  \alpha \to
0.
$$
Quantum states  (pure as well as mixed) are images of Gaussian
fluctuations of the magnitude $\alpha$ on the infinite dimensional
space $\Omega.$

In [1] we considered the quantum model based on the real Hilbert
space $H.$ This model is essentially simpler than the complex QM. It
is well know that justification of introduction  of the complex
structure in QM is a very complicated problem.
 We  show that the complex structure is the image of the
symplectic structure on the infinite dimensional phase space.

We found the classical Hamiltonian dynamics on the phase space which
induces the quantum state dynamics (Schr\"odinger's equation). The
crucial point is that the classical Hamilton function ${\cal
H}(\psi)$ should be $J$-{\it invariant:}
\begin{equation}
\label{SI} {\cal H}(J\psi)= {\cal H}(\psi),
\end{equation}
where $\psi \in \Omega= Q\times P, \; Q=P=H,$ and $J: Q\times P \to
Q\times P$ is the symplectic operator.  The main reason to consider
classical dynamics with $J$-invariant Hamilton functions is that
such dynamics preserve the magnitude of classical random
fluctuations: the dispersion of a Gaussian measure. In our approach
the conventional (linear) quantum dynamics is the image of the
classical  dynamics for a special class of quadratic Hamilton
functions, namely, satisfying the condition (\ref{SI}). Thus any
quantum dynamics is in fact dynamics of a classical (but
infinite-dimensional harmonic oscillator). Since all models under
consideration are statistical, dynamics of a quantum state
(including a pure state) is dynamics of {\it a Gaussian ensemble of
infinite-dimensional harmonic oscillators. For nonquadratic Hamilton
function classical dynamics on $\Omega$ can be represented as a
nonlinear Schr\"odinger equation.} Thus by representing QM as the
image of PCSFT we see that the nonlinear Schr\"odinger equation is
not less natural than the conventional linear equation.

%Our approach is based on two scalings of the classical prequantum
%model based on two small parameters $h > 0$ and $\alpha > 0.$ Here
%$h$ is the time-scaling parameter. The internal time of system
%(=field) evolution is given by the parameter $\tau \in {\bf R}$ and
%the time of observable is given by the parameter $t \in {\bf R}$ and
%these parameters are coupled through:functions
%\begin{equation}
%\label{SIT} \tau =\frac{t}{h}.
%\end{equation}
%Thus each unit interval of time for an observer corresponds to a
%huge interval of the internal time. In the physical model the
%parameter $h$ is the dimensionless parameter having the magnitude of
%the Planck constant.

Our approach is based on scaling of the classical prequantum model
based on a small parameters  $\alpha > 0.$ The parameter $\alpha$
 describes the magnitude
(dispersion) of quantum fluctuations. In our approach quantum
averages are obtained as approximations of classical averages (when
$\alpha\to 0)$ for amplified classical variables. If for a classical
variable $f(\psi)$ we define its {\it amplification} by
$$
f_\alpha(\psi)= \frac{1}{\alpha} f(\psi)
$$
then (\ref{AVP}) implies that
\begin{equation}
\label{AVP0} <f>_{\rm{quantum}}= \lim_{\alpha\to 0}
<f_\alpha>_{\rm{classical}}
\end{equation}
In the first version of our approach [1] we identified the parameter
$\alpha$  with the Planck constant $h$ (all parameters were
considered as dimensionless). This was motivated by SED and Nelson's
stochastic QM in that quantum fluctuations have the Planck
magnitude. However, our own model does not say anything about
relation of the Planck constant and the magnitude of quantum
fluctuations. We could not exclude the possibility that the
$\alpha$-scale is essentially finer that the SED-scale based on the
Planck constant $h.$

Thus in our approach QM is a theory about {\it amplification of
quantum fluctuations,} fluctuations of the prequantum classical
field (``background field'').\footnote{It is a good place to cite a
remark of Greg Jaeger at the round table of the conference QTRF-3
(V\"axj\"o-2005): ``Quantum fluctuations are very important. We
actually amplify them in our laboratories using Parametric Down
Conversion,'' see [7].} So we are in the same camp with SED-people
with the only possible difference: the energy scale.

 We pay attention that any point wise classical dynamics (in particular, Hamiltonian) can be lifted
to spaces of variables (functions) and statistical states
(probability measures). In the case of a $J$-invariant Hamilton
function by mapping these lifting to QM we obtain, respectively,
Heisenberg's dynamics for quantum observables and von Neumann's
dynamics for statistical operators.

We emphasize that one should distinguish (as always in classical
statistical physics) dynamics of states of individual physical
systems (point wise dynamics) and dynamics of statistical states
(dynamics of probability distributions). In conventional QM  these
two dynamics are typically identified. Our approach supports the
original views of E. Schr\"odinger [8], [9].  Schr\"odinger's
equation is a special type of the Hamiltonian equation on the
infinite-dimensional phase-space (the space of classical fields). By
our interpretation this equation describes the evolution of
classical states (fields). It is impossible to provide any
statistical interpretation to such individual states. In particular,
the wave function considered as a field satisfying Schr\"odinger's
equation has no statistical interpretation. Only statistical states
(probability measures in the classical model) and corresponding
density operators (which are in fact scalings of covariation
operators of measures representing statistical states) have a
statistical interpretation. The root of misunderstanding was
assigning (by M. Born) the statistical interpretation to the wave
function and at the same time considering it as the complete
description of an individual quantum system (the Copenhagen
interpretation). The tricky thing is that in fact Born's
interpretation should be assigned not to an individual state $\Psi,$
but to a statistical state given by the Gaussian distribution with
the covariation operator:
\begin{equation}
\label{PSI} B_\Psi = \alpha \; \Psi \otimes \Psi.
\end{equation}
Thus pure quantum states are simply statistical mixtures of special
Gaussian fluctuations (concentrated on two dimensional (real)
subspaces of the infinite dimensional Hilbert space), see section 9
for details. One could reproduce dynamics of such a statistical
state by considering the Schr\"odinger equation with random initial
conditions:
\begin{equation}
\label{SRI} i h \frac{d\xi}{d t}(t;\psi) = {\bf H} \xi(t;\psi),
\xi(t_0;\psi)= \xi_0(\psi),
\end{equation}
where ${\bf H}$ is Hamiltonian and $\xi_0(\psi)$ is the initial
Gaussian random vector taking values in  the Hilbert space. We
emphasize that $\Vert \xi(t;\psi) \Vert\in [0, +\infty).$ There is
no place for the standard normalization condition: $\Vert
\xi(t;\psi) \Vert=1.$ Quantum randomness is not irreducible, cf.
with von Neumann [10]. This is classical randomness of initial
conditions, cf. with Bohmian mechanics [11], [12].

We remark that the idea that QM can be represented as a
probabilistic projection of a classical probabilistic model was
elaborated in the series of author's papers, see, e.g., [13]-[17].
In these papers there was introduced {\it prespace} in that it is
possible to provide  a finer description of complexes of physical
conditions (physical contexts) than in QM. Quantum states were
obtained as images of contexts. In the present paper the role of
prespace is played by the phase space $\Omega,$ contexts are
represented by special Gaussian ensembles of classical fields.

Finally, we pay attention that our work might be considered as a
realization of ``Einstein's dream'': creation of purely field model,
cf. [18], [19].

\section{Hamiltonian  mechanics}

\subsection{Quadratic Hamilton function}

We consider the conventional classical phase space:
$$
\Om=Q \times P, \; \; Q= P = \br^n
$$
Here states are represented by points $\psi= (q, p) \in \Om;$
evolution of a state is described by the Hamiltonian equations
\begin{equation}
\label{HE} \dot q = \frac{\partial {\cal H}}{\partial p},\; \;  \dot
p=-\frac{\partial {\cal H}}{\partial q,}
\end{equation}where ${\cal H}(q, p)$ is the Hamilton function (a real valued function on the phase space $\Om).$

We consider the scalar product on $\br^n:$ $ (x, y)=\sum_{j=1}^n x_j
y_j $ and define the scalar product on $\Omega:$ $(\psi_1,
\psi_2)=(q_1, q_2) +(p_1, p_2).$ In our reseach we shall be
interested in quadratic Hamilton functions:
\begin{equation}
\label{V} {\cal H}(q, p)=\frac{1}{2} ({\bf H} \psi,\psi),
\end{equation}where ${\bf H}: \Om \to \Om$ is a symmetric operator.
We remark that any ($\br$-linear) operator $A: {\bf R}^{2n} \to {\bf R}^{2n}$ can be represented in the form
\[A= \left( \begin{array}{ll}
A_{11}&A_{12}\\
A_{21}&A_{22}
\end{array}
\right ),
\]
where $A_{11}: Q \to Q, A_{12}: P \to Q,$ $A_{21}: Q \to Q, A_{22}:
P \to P.$ A linear operator $A: {\bf R}^{2n} \to {\bf R}^{2n}$ is
symmetric if $$A_{11}^*=A_{11},\; A_{22}^*=A_{22},\;
A_{12}^*=A_{21},\; A_{21}^*=A_{12}.
$$
Thus the Hamilton function (\ref{V}) can be written as:
\begin{equation}
\label{V0}
{\cal H}(q,p) = \frac{1}{2} [({\bf H}_{11} q, q)+2({\bf H}_{12} p, q) + ({\bf H}_{22} p,p)],
\end{equation}
The Hamiltonian equation  is linear  and it has the form:
\begin{equation}
\label{HE20}
\dot q= {\bf H}_{21}q + {\bf H}_{22} p,  \; \;   \dot p=-( {\bf H}_{11}q +{\bf H}_{12}p)
\end{equation}
As always,  we define the canonical symplectic structure  on the
phase space $\Omega=Q\times P$ starting with the {\it symplectic
operator}
 \[J= \left( \begin{array}{ll}
 0&1\\
 -1&0
 \end{array}
 \right )
 \]
(here the blocks "$\pm 1$" denote $n \times n$ matrices with $\pm 1$ on the diagonal).
By using the symplectic operator $J$ we can write these Hamiltonian equations in the operator form:
\begin{equation}
\label{Y} \dot \psi= \left( \begin{array}{ll}
\dot q\\
\dot p
\end{array}
\right )=J{\bf H} \psi
\end{equation}
Thus
\begin{equation}
\label{Y2} \psi(t)= U_t \psi, \; \; \mbox{where} \; U_t=e^{J {\bf H}
t}.
\end{equation}
The map $U_t\psi$ is a linear Hamiltonian flow on the phase space
$\Omega.$

\subsection{$J$-invariant quadratic forms and
$J$-commuting operators} In our investigations we shall be
concentrated on consideration of $J$-invariant quadratic forms. It
is easy to see that symplectic invariance of the quadratic form
$f_A(\psi)= (A\psi, \psi)$, where $A: \Omega \to \Omega$ is the
linear symmetric operator,  is equivalent to commuting of $A$ with
the symplectic operator $J.$ Let us consider the class ${\cal
L}_{\rm symp} \equiv {\cal L}_{\rm symp} (\Omega)$ of  linear
operators $A: \Omega \to \Omega$ which commute with the symplectic
operator:
\begin{equation}
\label{SS}
A J= J A
\end{equation}
This is a subalgebra of the algebra of all linear operators ${\cal
L}(\Omega).$ We call such operators $J$-commuting.

{\bf Proposition 2.1.} {\it $A\in {\cal L}_{\rm symp} (\Omega)$ iff
$A_{11}=A_{22}=D, A_{12}=-A_{21}=S:$}
\[A= \left( \begin{array}{ll}
D&S\\
-S&D
\end{array}
\right )
\]

We remark that an operator $A \in {\cal L}_{\rm symp} (\Omega)$ is
symmetric iff $D^*=D$ and $S^*=-S.$  Hence any symmetric
$J$-commuting operator in the phase space is determined by a pair of
operators $(D, S)$, where $D$ is symmetric and $S$ is
anti-symmetric. Such an operator induces the quadratic form
\begin{equation}
\label{QF} f_A(\psi)=(A\psi, \psi)=(Dq, q)+2(Sp, q)+(Dp, p).
\end{equation}

\subsection{Dynamics for $J$-invariant quadratic Hamilton
functions} Let us consider an  operator  ${\bf H} \in {\cal L}_{\rm
symp}(\Omega)$: ${\bf H}= \left( \begin{array}{ll}
R&T\\
-T&R
\end{array}
\right ).$ This operator defines the quadratic  Hamiltonian function
${\cal H}(q, p)=\frac{1}{2} ({\bf H} \psi,\psi)$ which can be
written as
\begin{equation}
\label{HE1}{\cal H}(q, p)=\frac{1}{2}[(R p, p) + 2 (Tp, q) + (Rq, q)]
\end{equation}
where
$R^*=R , \; \; T^*=-T$
Corresponding Hamiltonian equations have the form
\begin{equation}
\label{HE2}
\dot q=Rp-Tq, \; \;  \dot p=-(Rq + Tp)
\end{equation}

{\bf Proposition 2.2.} {\it For a  $J$-invariant Hamilton function,
the Hamiltonian flow $U_t,$ see (\ref{Y2}), consists of
$J$-commuting operators:} $ U_t J=J U_t. $

{\bf Example 2.1.} (One dimensional $J$-invariant harmonic
oscillator) Let ${\cal H}(q, p)=\frac{1}{2} [\frac{p^2}{m} + m k^2
q^2]$ (we use the symbol $k$ to denote frequency, since $\psi$ is
already used for the point of the phase space).
 To get a Hamiltonian of the form (\ref{HE1}), we consider the case $\frac{1}{m}=m k^2.$ Thus
$\label{HE3}m=\frac{1}{k}$
and
%\begin{equation}
%\label{HE30}
${\cal H}(q, p)=\frac{k}{2} [ p^2 + q^2] ;$
%\end{equation}
Hamiltonian equations are given by
%\begin{equation}
%\label{HE4}
$ \dot q=k p,  \; \;  \dot p =-k q$
%\end{equation}
Here the symmetric $J$-commuting matrix ${\bf H}= \left(
\begin{array}{ll}
k&0\\
0&k
\end{array}
\right).$

Let us define the {\it symplectic form} on the phase space:
\begin{equation}
\label{SF} w(\psi_1, \psi_2)=(\psi_1, J\psi_2).
\end{equation}
Thus
$$
w(\psi_1, \psi_2)=(p_2, q_1)-(p_1, q_2)
$$
for $\psi_j= \{ q_j, p_j \}, j=1,2.$ This is a {\it skew-symmetric
bilinear form.}

{\bf Proposition 2.3.} {\it Let $A$ be a symmetric operator. Then $A \in {\cal L}_{\rm symp}(\Omega)$ iff it is symmeric with respect to the symplectic form:}
\begin{equation}
\label{0} w(A \psi_1, \psi_2)=w(\psi_1, A\psi_2)
\end{equation}

{\bf Remark 2.1.} (K\"aller structure) We started our considerations
not directly with an arbitrary symplectic form on $\Omega,$  but
with the canonical symplectic form (\ref{SF}) corresponding to the
inner product (Riemannian metric) on $\Omega.$ Thus we can
canonically introduce the {\it hermitian metric} on the complex
realization $\Omega_c$ of $\Omega.$ Thus, in fact, from the very
beginning we worked not on an arbitrary symplectic manifold, but on
a {\it K\"aller manifold.} The $J$-invariance appears very naturally
as the consistency condition for the Riemannian metric and the
symplectic structure.

\subsection{Complex representation of dynamics for $J$-invariant
quadratic Hamilton functions}

Let us introduce on phase space $\Omega$ the complex structure:
$\Om_c=Q \oplus i P.$ We have $i\psi=-p + iq=- J\psi.$ A
$\br$-linear operator $A:\Om_c \to \Om_c$ is ${\bf C}$-linear iff
$A(i\psi)=iA\psi$ that is equivalent to $A \in {\cal L}_{\rm
symp}(\Omega).$

\medskip

{\bf Proposition 2.4.} {\it The class of ${\bf C}$-linear operators
${\cal L} (\Om_c)$ coincides with the class of $J$-commuting
operators ${\cal L}_{\rm symp}(\Om).$}

\medskip

We introduce on $\Om_c$ a complex scalar product (hermitian metric,
see Remark 2.1) based on the ${\bf C}$-extension of the real scalar
product:
$$
<\psi_1, \psi_2>=<q_1 + ip_1, q_2 + ip_2>
$$
$$
=(q_1, q_2) + (p_1, p_2) + i((p_1, q_2) - (p_2, q_1)).
$$
Thus
$$
<\psi_1, \psi_2>=(\psi_1, \psi_2)-i w(\psi_1, \psi_2),
$$
where $w$ is the symplectic form. This is the canonical hermitian
metric on the K\"aller manifold $\Omega.$

A ${\bf C}$-linear operator $A$  is symmetric with respect to the
complex scalar product $<\ldots>$ iff it is symmetric with respect
to both real bilinear forms: ($\cdot, \cdot$) and $w(\cdot, \cdot)$.
Since for $A \in  {{\cal L}_{\rm symp}}(\Omega)$ the former implies
the latter,  we get that a ${\bf C}$-linear operator in $\Omega_c$is
symmetric iff it is symmetric in the real space $\Omega.$

\medskip

{\bf Proposition 2.5.} {\it The class of ${\bf C}$-linear symmetric
operators ${\cal L}_{\rm{s}}(\Omega_c)$ coincides with the class of
$J$-commuting symmetric operators ${\cal L}_{\rm symp,s}(\Omega).$}

\medskip

We also remark that for a $J$-commuting operator $A$ its {\it real
and complex adjoint operators:}
$$
A^\star\; \mbox{and} \; A^*
$$
coincide. We showed
that ${\bf C}$-linear symmetric operators appear naturally as
complex representations of $J$-commuting symmetric operators.

\medskip

{\bf Proposition 2.6.} {\it For a quadratic $J$-invariant Hamilton
function ${\cal H}(\psi),$ its complexification does not change
dynamics.}

{\bf Proof.} To prove this, we remark that $w( {\bf H} \psi,
\psi)=0$ and hence
$$
{\cal H}(\psi)= \frac{1}{2} <{\bf H} \psi, \psi> =\frac{1}{2} [({\bf
H}\psi, \psi)- i w({\bf H}\psi, \psi)]= \frac{1}{2} ({\bf H}\psi,
\psi) , \psi \in \Om.
$$

I consider the introduction of a complex structure on the
phase-space merely  as using a new language: instead of symplectic
invariance, we speak about ${\bf C}$-linearity. By Proposition 2.6
the Hamilton function (\ref{HE1}) can be written ${\cal
H}(\psi)=\frac{1}{2} <{\bf H}\psi, \psi>, {\bf H} \in {\cal
L}_{\rm{s}}({\bf C}^{n}),$ and the  Hamiltonian equation (\ref{Y})
can be written in the complex form as:
\begin{equation}
\label{Y10} i \frac{d \psi}{d t} = {\bf H} \psi
\end{equation}
Any solution has the following complex representation:
\begin{equation}
\label{Y20} \psi(t)=U_t \psi, \; \; U_t=e^{-i{\bf H} t/h}.
\end{equation}
This is the complex representation of flows corresponding to
quadratic $J$-invariant Hamilton functions.

\section{Schr\"odinger dynamics as a dynamics with $J$-invariant Hamilton function on
the infinite dimensional phase space}

Let $\Om_c$ be a complex Hilbert space (infinite dimensional and
separable) and let $<\cdot, \cdot>$ be the complex scalar product on
$\Om_c.$ The symbol ${\cal L}_{\rm{s}}\equiv {\cal
L}_{\rm{s}}(\Omega_c)$ denotes the space of continuous
 ${\bf C}$-linear self-adjoint operators. We use the Planck system of units: $h=1.$
The Schr\"odinger dynamics in $\Om$ is given by the linear equation:
\begin{equation}
\label{S} i  \frac{d \psi}{d t} = {\bf H} \psi
\end{equation}
and hence
\begin{equation}
\label{S0} \psi(t)=U_t \psi, \; \; U_t=e^{-i{\bf H}t}.
\end{equation}
 We see that these are
simply infinite-dimensional versions of equations (\ref{Y10}) and
$(\ref{Y20})$ obtained from the Hamiltonian equations for a
quadratic $J$-invariant Hamilton function in the process of
complexification of classical mechanics. Therefore we can reverse
our previous considerations (with the only remark that now the phase
space is infinite dimensional) and represent the Schr\"odinger
dynamics (\ref{S}) in the complex Hilbert space as the Hamiltonian
dynamics in the infinite-dimensional phase space.\footnote{Infinite
dimension induces merely mathematical difficulties. The physical
interpretation of formalism is the same as in the finite-dimensional
case.} We emphasize that this Hamiltonian dynamics (\ref{Y}) is a
dynamics in the phase space $\Omega$ and not in the unit sphere of
this Hilbert space! The Hamiltonian flow $\psi(t, \psi)=U_t\psi$ is
a flow on the whole phase space $\Omega.$

We consider in $\Om$ the ${\bf{R}}$-linear operator $J$ corresponding to
multiplication by $-i;$ we represent the complex Hilbert space in the form:
$$
\Om_c=Q \oplus i P,
$$
where $Q$ and $P$ are copies of the real Hilbert space. Here
$\psi=q+ip.$ We emphasize that $q$ and $p$ are not ordinary position
and momentum for particles. These are their field analogues (if we
choose $Q=P=L_2({\bf R}^3)$): these are functions of $x\in {\bf
R}^3.$ We consider now the real phase space:
$$
\Om=Q \times P.
$$
As in the finite dimensional case, we have:

\medskip

{\bf Proposition 3.1.} {\it The class of continuous ${\bf C}$-linear
self-adjoint operators ${\cal L}_{\rm{s}}(\Omega_c)$ coincides with
the class of  continuous $J$-commuting self-adjoint operators ${\cal
L}_{\rm symp,s}(\Omega).$}

\medskip

Let us consider a quantum Hamiltonian ${\bf H} \in {\cal
L}_{\rm{s}}(\Omega_c)$.\footnote{We may consider operator ${\bf H}
\geq 0,$ but for the present consideration this is not important.}
It determines the classical Hamiltonian function:
\[{\cal H}(\psi)=\frac{1}{2}  <{\bf H} \psi, \psi>= \frac{1}{2} [(Rp, p) + 2(Tp, q)+(Rq, q)]\]
The corresponding Hamiltonian equation on the classical phase space
$\Om=Q\times P,$ where $Q$ and $P$ are copies of the real Hilbert
space, is given by
\begin{equation}
\label{Z}  \frac{d q}{dt} = Rp-Tq, \; \frac{d p}{dt}=-(Rq + Tp)
\end{equation}
If we apply the complexification procedure to this system of Hamiltonian equations we,
of course, obtain the Schr\"odinger equation (\ref{S}).

{\bf Example 3.1.} Let us consider an important class of Hamilton
functions
\begin{equation}
\label{HF} {\cal H} (q, p)=\frac{1}{2}[(Rp, p)+(Rq, q)],
\end{equation}
where $R$ is a symmetric operator. The corresponding Hamiltonian
equations have the form:
\begin{equation}
\label{HF1} \dot q=Rp, \; \dot p=-Rq.
\end{equation}
We now choose $H=L_2(\br^3),$ so $q(x)$ and $p(x)$ are components of
the vector-field $\psi(x)=(q(x), p(x)).$ We can call fields $q(x)$
and $p(x)$ {\it mutually inducing.} The presence of the field $p(x)$
induces dynamics of the field $q(x)$ and vice versa, cf. with
electric and magnetic components, $q(x)=E(x)$ and $p(x)=B(x),$ of
the classical electromagnetic field, cf. Einstein and Infeld [19],
p. 148: {\small ``Every change of an electric field produces a
magnetic field; every change of this magnetic field produces an
electric field;  every change of ..., and so on.''} We can write the
form (\ref{HF}) as
\begin{equation}
\label{HF2} {\cal H} (q, p)=\frac{1}{2} \int_{\br^6} R(x, y) [q (x)
q(y) + p(x) p(y)] dx dy
\end{equation}
or
\begin{equation}
\label{HF3} {\cal H}(\psi)=\frac{1}{2} \int_{\br^6} R(x, y) \psi (x)
\bar{\psi} (y) dx dy ,
\end{equation}
where $R (x, y)=R(y, x)$ is in general a distribution on $\br^6.$

We call such a  kernel $R(x, y)$  a {\it self-interaction potential}
for the field $\psi(x)=(q, (x), p(x)).$ We pay attention that $R(x,
y)$ induces a self-interaction of each component of the $\psi(x),$
but there is no cross-interaction between components $q(x)$ and
$p(x)$ of the vector-field $\psi(x).$

\medskip

One may justify consideration of $J$-invariant physical variables on
the Hilbert phase space by referring to quantum mechanics: ``the
correct classical Hamiltonian dynamics is based on $J$-invariant
Hamilton functions, because they induce the correct quantum
dynamics.'' So the classical prequantum dynamics was reconstructed
on the basis of the quantum dynamics. I have nothing against such an
approach. But it would be interesting to find internal classical
motivation for considering $J$-invariant Hamilton functions. We
shall do this in section 5.

\section{Lifting of point wise dynamics to spaces of variables and measures}

\subsection{General dynamical framework}
Let $(X, F)$ be an arbitrary measurable space. Here $X$ is a set and
$F$ is a $\sigma$-field of its subsets. Denote the space of random
variables (measurable maps $f: X\to {\bf R})$ by the symbol $RV(X)$
and the space of probability measures on $(X, F)$ by the symbol
$PM(X).$ Consider a measurable map $g:X \to X.$ It induces the maps
$$
g^*: RV(X) \to RV(X), \; \; g^* f(x) =f(g(x))
$$
$$
g^*: MP(X) \to MP(X),  \; \;  \int_X f(x) dg^* \mu(x)  = \int_X g^*
f(x) d \mu(x).
$$
We now consider a dynamical system in $X:$
\begin{equation}
\label{PW0}
x_t= g_t(x),
\end{equation}
 where $g_t:X \to X$ is an one-parametric family of measurable maps
(the parameter $t$ is real and plays the role of time). By using lifting $\alpha$ and $\beta$ we can lift
this point wise dynamics in $X$ to dynamics in $RV(X)$ and $MP(X),$  respectively:
\begin{equation}
\label{PW}f_t= g^*_t f
\end{equation}
\begin{equation}
\label{PW1}\mu_t= g^*_t \mu.
\end{equation}
We shall see in sections 6, 7 that for $X= \Omega$ (infinite
dimensional phase space)  quantum images of dynamical systems
(\ref{PW0}), (\ref{PW}), (\ref{PW1}) are respectively dynamics of
Schr\"odinger (for state -- wave function), Heisenberg (for
operators-observables) and von Neumann (for density operator). To
obtain quantum mechanics, we should choose adequate spaces of
physical variables and measures.

\subsection{Lifting of the Hamiltonian dynamics} It is well known that the
lifting of Hamiltonian dynamics to the space of smooth variables is
given by the {\it Liouville equation}, see e.g. [20]. In particular,
the functional lifting of any Hamiltonian dynamics on the Hilbert
phase space $\Omega$ can be represented as the infinite-dimensional
Liouville equation, [21]. We remark that this is a general fact
which has no relation to our special classical framework based on
$J$-invariant Hamilton functions. For smooth functions on the
$\Omega$ we introduce the Poisson brackets, see, e.g., [22]:
$$
\{ f_1(\psi), f_2(\psi)\}= \Big(\frac{\partial f_1}{\partial
q}(\psi), \frac{\partial f_2}{\partial p}(\psi)\Big) -
\Big(\frac{\partial f_2}{\partial q}(\psi), \frac{\partial
f_1}{\partial p}(\psi)\Big).
$$
We recall that for $f:H \to {\bf R}$ its first derivative can be
represented as a vector belonging $H;$ so for $f: H \times H\to {\bf
R}$ its gradient $\nabla f(\psi)$ belongs $H \times H.$ We pay
attention that $\{f_1, f_2\}= =(\nabla f_1, J\nabla f_2)= w(\nabla
f_1, \nabla f_2).$ Let ${\cal H}(\psi)$ be a smooth Hamilton
function inducing the flow $U_t(\psi).$ For a smooth function $f_0$
we set $f(t, \psi)= f_0(U_t(\psi)).$ It is easy to see that this
function is the solution of the Cauchy problem for the Liouville
equation:
\begin{equation}
\label{PWZ}\frac{\partial f}{\partial t}(t,\psi)= \{ f(t, \psi),
{\cal H} (\psi)\}, \; f(0, \psi)=f_0(\psi)
\end{equation}
The functional flow $\Psi(t,f_0)= \alpha_{U_t} f_0$ can be represented as
\begin{equation}
\label{PWZA}
\Psi(t,f_0)= e^{-t L} f_0,
\end{equation}
where
$$
L=\Big(\frac{\partial {\cal H} }{\partial q}(\psi), \frac{\partial
}{\partial p}\Big) - \Big(\frac{\partial {\cal H} }{\partial
p}(\psi), \frac{\partial }{\partial q} \Big)
$$

\section{Dispersion preserving dynamics of statistical states}

Everywhere in this section we consider only quadratic Hamilton
functions on the infinite-dimensional phase space $\Omega.$ We start
our consideration with an arbitrary quadratic Hamiltonian function
${\cal H}(\psi)= \frac{1}{2} ({\bf H} \psi, \psi)$ (the operator
${\bf H}$ need not be $J$-commuting). Let us consider the
Hamiltonian flow $U_t: \Omega \to \Omega$ induced by the Hamiltonian
system (\ref{Y}). This map is given by (\ref{Y2}). It is important
to pay attention that the map $U_t$ is invertible; in particular,
\begin{equation}
\label{PWO}U_t(\Omega)= \Omega.
\end{equation}We are interested in a Hamiltonian flow $U_t$ such that the corresponding dynamics in the space of probabilities
(\ref{PW1}) preserves the magnitude of statistical fluctuations:
\begin{equation}
\label{PWD} \sigma^2(U_t^* \rho) =\sigma^2(\rho): \int_{\Omega}
\Vert \psi \Vert^2 d U_t^* \rho(\psi) = \int_{\Omega} \Vert \psi
\Vert^2 d \rho(\psi)
\end{equation}
or
\begin{equation}
\label{PWD0} \int_{\Omega} \Vert U_t\psi \Vert^2 d\rho (\psi)=
\int_{\Omega} \Vert \psi \Vert^2 d\rho (\psi) .
\end{equation}
We start the study of this problem with a sufficient condition for
preserving the magnitude of statistical fluctuations: the
Hamiltonian flow $U_t\psi$ consists of isometric maps:
\begin{equation} \label{PWN} \Vert U_t \psi \Vert^2 = \Vert \psi
\Vert^2, \psi \in \Omega.
\end{equation}

{\bf Proposition 5.1.} {\it  A Hamiltonian flow $U_t\psi$ is
isometric iff the function  ${\cal H}(\psi)$ is $J$-invariant.}

{\bf Proof.} a). Let ${\cal H}$ be $J$-commuting. Then we have:
$$
\frac{d}{d t} \Vert U_t \psi\Vert^2= 2  (\dot{U_t}\psi, U_t \psi) =
2(J {\bf H} U_t \psi, U_t \psi)=0
$$
Here we used the simple fact that the operator $J{\bf H}$ is  skew symmetric: $(J {\bf H})^\star= - {\bf H}J= - J{\bf H}.$
Thus (\ref{PWN}) holds.

b). Let  (\ref{PWN}) hold. Then $\frac{d}{d t} \Vert U_t \psi
\Vert^2=0.$ By using previous computations and (\ref{PWO}) we get
that: \begin{equation} \label{PWNS} (J {\bf H} \psi, \psi)=0, \;
\psi \in \Omega.
\end{equation}
Hence the operator $J{\bf H}$ is  skew symmetric. This implies that
${\bf H}$ commutes with $J.$

\medskip

For our further considerations, see section 12, it is useful to
rewrite (\ref{PWNS}) in the form:
\begin{equation}
\label{PWNS1} (J {\cal H}^{\prime}(\psi), \; \psi)=0, \psi \in
\Omega.
\end{equation}

{\bf Corollary 5.1.} {\it  The flow corresponding to a $J$-invariant
Hamilton function preserves the fluctuations of the fixed
$\alpha$-magnitude. For any measure $\rho$ (having the zero mean
value and finite dispersion), if $\sigma^2(\rho)=\alpha,$ then
$\sigma^2(U_t^* \rho)=\alpha$ for any $t\geq 0.$}

\medskip

This is our explanation of the exceptional role of $J$-invariant
physical variables on the infinite-dimensional classical phase
space.

If a Hamilton function is not $J$-invariant then the corresponding
Hamiltonian flow can induce increasing of the magnitude of
fluctuations. But we recall that quantum model is a representation
based on neglecting by fluctuations of the magnitude $o(\alpha),
\alpha\to 0.$ Therefore  a Hamiltonian flow which is not
$J$-invariant can induce the transformation of ``quantum statistical
states'', i.e., distributions on the phase space having dispersion
of the magnitude $\alpha,$ into ``nonquantum statistical states'',
i.e. distributions on the phase space having dispersions essentially
larger than $\alpha.$

\section{Dynamics in the space of  physical variables}

\subsection{Arbitrary quadratic
variables} Let us consider the  Hamiltonian flow $U_t:  \Omega \to
\Omega$ induced by an arbitrary  quadratic Hamilton function. Let
$A:\Omega \to \Omega$ be a continuous self-adjoint operator and
$f_A=(A\psi, \psi).$ We have $U_t^* f_A(\psi)= f_A(U_t \psi)=
f_{U_t^\star A U_t}(\psi).$ This dynamics can be represented as the
dynamics in the space of continuous linear symmetric operators
\begin{equation}
\label{DH}A_t= U_t^\star A U_t
\end{equation}
We remark that $U_t= e^{J{\bf H}t},$ so $U_t^\star= e^{-{\bf H}J
t}.$ Thus
\begin{equation}
\label{DHZ}A_t= e^{-{\bf H}J t} A e^{J{\bf H}t}.
\end{equation}Thus
$\frac{d A_t}{d t} =  \Big( A_t J{\bf H} - {\bf H}J A_t\Big),$ or
\begin{equation}
\label{DH0} \frac{d A_t}{d t} = [A_t, {\bf H}J]+  A_t [J, {\bf H}]
\end{equation}
We remark that dynamics (\ref{DH}) can be also obtained from the
Liouville equation, but I presented the direct derivation.

\subsection{$J$-invariant variables} We consider the
space of physical variables
$$
V_{\rm{quad, symp}}(\Omega)=\{ f: \Omega \to {\bf R}: f \equiv
f_A(\psi) = \frac{1}{2} (A \psi, \psi), A \in {\cal L}_{\rm{symp},
s}(\Omega) \}
$$
(consisting of $J$-invariant quadratic forms).  Let us consider the
lifting of the  flow corresponding to a $J$-invariant quadratic
Hamilton function to the space $V_{\rm{quad, symp}}(\Omega).$ In
this case both operators, ${\bf H}$ and $A$ are $J$-commuting.
Therefore the flow (\ref{DHZ}) can be written as
\begin{equation}
\label{DHZQ}A_t= U_t^\star A U_t = e^{-J {\bf H} t} A e^{J{\bf H}t}
\end{equation}
The evolution equation (\ref{DH0}) is simplified:
\begin{equation}
\label{DH0A} \frac{d A_t}{d t} = -J[ {\bf H},A_t]
\end{equation}

\subsection{Complexification} As in section 6.2, we suppose that
$[{\bf H}, J]=0$ and $[A, J]=0.$ By considering on the phase space
the complex structure and representing the symplectic operator  $J$
by $-i$ we write (\ref{DHZ}) in the form of the Heisenberg dynamics:
\begin{equation}
\label{DHG} A_t= U_t^* A U_t=e^{it{\bf H}} A e^{-it{\bf H}}
\end{equation}
(here $U_t^*$ is the complex adjoint operator to $U_t)$
and the evolution equation (\ref{DH0}) in the form of the Heisenberg equation:
\begin{equation}
\label{DH0G} \frac{d A_t}{d t} = i [ {\bf H},A_t]
\end{equation}
Thus this equation is just the image of the lifting of the classical
quadratic Hamiltonian dynamics  in the case of $J$-invariant
quadratic variables.

\section{Dynamics in the space of statistical states}

\subsection{Arbitrary Gaussian
measures} Let us consider the flow $U_t:  \Omega \to \Omega$ induced
by an arbitrary quadratic Hamilton function ${\cal H}(\psi).$  Let
$\rho$ be an arbitrary Gaussian measure with zero mean value. Since
a linear continuous transformation of a Gaussian measure is again a
Gaussian measure, we have that $U_t^*(\rho)$ is Gaussian. We find
dynamics of the covariation operator of $U_t^*(\rho).$ We have:
$$
(\rm{cov}(U_t^*\rho) y_1, y_2)=   \int_\Om  (y_1, \psi) (y_2, \psi)
d U_t^* \rho (\psi)
$$
$$
= \int_\Om  (y_1, U_t \psi) (y_2, U_t\psi) d  \rho (\psi)= (\rm{cov}
(\rho) U_t^\star y_1, U_t^\star y_2).
$$
Thus, for the covariation operator $B_t= \rm{cov}(U_t^*\rho),$ we
have:
\begin{equation}
\label{DH0M} B_t= U_t B U_t^\star \equiv e^{J{\bf H} t} B e^{-{\bf
H}J t}
\end{equation}
Thus $\frac{d B_t}{d t} = \Big( J{\bf H}B_t  -  B_t{\bf H}J\Big),$
or
\begin{equation}
\label{DHM} \frac{d B_t}{d t} = [J{\bf H}, B_t]+ B_t [J, {\bf H}]
\end{equation}

\subsection{$J$-invariant measures} We now consider the
lifting (to the space of measures)  of the flow $U_t:  \Omega \to
\Omega$ induced by a $J$-invariant quadratic Hamilton function
${\cal H}(\psi).$ We start with the following mathematical result:

{\bf Proposition 7.1.} {\it A Gaussian measure $\rho$ (with zero
mean value) is $J$-invariant   iff  its covariation operator is
$J$-invariant.}

{\bf Proof.} a). Let $J^* \rho= \rho.$ It is sufficient to prove
that $B J$ is skew symmetric, where $B=\rm{cov} \; \rho.$ We have:
$$
(BJy_1, y_2)= \int_\Omega (Jy_1, \psi) (y_2, \psi) d \rho(\psi)= -
\int_\Omega (y_1, J\psi) (y_2,  J^\star J \psi) d \rho(\psi)
$$
$$
=- \int_\Omega (y_1, J\psi) (Jy_2,   J \psi) d \rho(\psi)= -
\int_\Omega (y_1, \psi) (Jy_2,   \psi) d \beta_J \rho(\psi)
$$
$$
= - \int_\Omega (Jy_2,   \psi) (y_1, \psi) d \rho(\psi)=
-(BJy_2,y_1)= -(y_1, BJy_2).
$$
b). Let $B=\rm{cov}(\rho)\in {\cal L}_{\rm{symp}, s}(\Omega).$ We
find the Fourier transform of the Gaussian measure $J^* \rho:$
$$
\widetilde{J^* \rho}(y)= \int_\Omega e^{i(y, J\psi)} d \rho (\psi)=
=\tilde{\rho}(J^\star y)= e^{-\frac{1}{2} (B J^\star y, J^\star y)}=
=\tilde{\rho}(y).
$$

From the proof we also obtain:

{\bf Corollary 7.1.} {\it Let $\rho$ be an arbitrary $J$-invariant
measure. Then its covariation operator is $J$-invariant.}

Since the flow for a $J$-invariant quadratic Hamilton function
consists of $J$-commuting linear operators, $J U_t= U_t J,$ by using
the representation (\ref{DH0M}) and Proposition 7.1 we prove that
the space of $J$-invariant Gaussian measures (with zero mean value)
is invariant for the map $U_t^*.$  Here we have:
\begin{equation}
\label{DH0MA} B_t= U_t B U_t^\star \equiv e^{J{\bf H}t} B e^{-J{\bf
H} t}
\end{equation}
 or
\begin{equation}
\label{DHMA} \frac{d B_t}{d t} = -J [B_t, {\bf H}]
\end{equation}

\subsection{Complexification} We suppose that
$[{\bf H}, J]=0$ and $[B, J]=0.$ By considering on the phase space
the complex structure and representing the symplectic operator  $J$
by $-i$ we write (\ref{DH0MA}) in the form:
\begin{equation}
\label{DH0MB} B_t=U_t B U_t^*=  e^{-i{\bf H}t} B e^{i{\bf H} t}
\end{equation}
 or
\begin{equation}
\label{DHMX} \frac{d B_t}{d t} = i[B_t, {\bf H}]
\end{equation}
This is nothing else than the von Neumann equation for the statistical operator. The only difference is that
the covariance operator $B$ is not normalized. The normalization will come from the correspondence map $T$
projecting a prequantum classical statistical model onto QM, see section 8.

\subsection{Dynamics in the space of statistical states} First we consider the
space of all Gaussian measures having zero mean value and dispersion
$\alpha.$ We recall that here $\alpha$ is a small parameter
characterizing  fluctuations of energy of  the background field:
$$
\alpha = \int_{L_2({\bf R}^3) \times L_2({\bf R}^3)} \int_{{\bf
R}^3}(\vert q(x) \vert^2 + \vert p(x) \vert^2) d x \; d \rho(q, p).
$$
We do not provide dimension analysis in this paper. But the crucial
point is that elements of the phase space $\Omega= L_2({\bf R}^3)
\times L_2({\bf R}^3)$, ``wave functions'', are considered as
classical fields (as the classical electromagnetic field) and not as
`square roots of probabilities'' (cf. with the conventional Born's
interpretation of the wave function, but also cf. with the original
Schr\"odinger's interpretation).

Denote this space of such measures by the symbol
$S_G^\alpha(\Omega).$ These are Gaussian measures such that
$$
(y, m_\rho)= \int_\Omega (y, \psi) d\rho(\psi)=0, y \in \Omega, \;
\mbox{and}\;
 \sigma^2(\rho)= \int_\Omega \Vert \psi\Vert^2 d \rho(\psi)= \alpha
$$

For the flow $U_t$ corresponding to a $J$-invariant quadratic
Hamilton function, we have (see section 5)
$$
U_t^*: S_G^\alpha(\Omega) \to S_G^\alpha(\Omega)
$$
Denote the subspace of $S_G^\alpha(\Omega)$ consisting of
$J$-invariant measures by the symbol $S_{G,
\rm{symp}}^\alpha(\Omega).$ We also have:
$$
U_t^*: S_{G, \rm{symp}}^\alpha(\Omega) \to S_{G,
\rm{symp}}^\alpha(\Omega).
$$

\subsection{Complex covariation} Everywhere below we consider only
measures with finite dispersions. Let us introduce  {\it complex
average $m_\rho^c$ and covariance operator} $B^c\equiv \rm{cov}^c
\rho$ by setting:
\begin{equation}
\label{CCAV} <m_\rho^c, y>=  \int_\Omega <y, \psi> d \rho (\psi).
\end{equation}
\begin{equation}
\label{CCV} <B^c y_1, y_2>= \int_\Omega <y_1, \psi> <\psi, y_2> d
\rho (\psi).
\end{equation}

{\bf Proposition 7.2.} {\it Let $\rho$ be a  $J$-invariant measure.
Then}
\begin{equation}
\label{GSAV}m_\rho^c =0 \; \mbox{iff}\; \;  m_\rho=0.
\end{equation}
{\bf Proof.} Since $\rho$ is $J$-invariant, for any Borel function
$f: \Omega \to {\bf R},$ we have:
\begin{equation}
\label{GSA0}\int_\Omega f(\psi_q,\psi_p) d\rho(\psi_q,\psi_p)=
\int_\Omega f(\psi_p, -\psi_q) d\rho(\psi_q,\psi_p)
\end{equation}
Let $m_\rho=0.$ Then:
$$
0=\int_\Omega (y, \psi) d\rho(\psi) = \int_\Omega [(y_q, \psi_q) +
(y_p, \psi_p)] d\rho(\psi)
$$
$$
=\int_\Omega [(y_q, \psi_p) - (y_p, \psi_q)] d\rho(\psi) =
\int_\Omega w(y, \psi)d\rho(\psi),
$$
where $w$ is the symplectic form $\Omega.$ Hence the last integral
is also equal to zero. On the other hand, for the complex average we
have:
\begin{equation}
\label{GSAC} <y, m_\rho^c> = \int_\Omega (y, \psi) d\rho(\psi) - i
\int_\Omega w(y, \psi)d\rho(\psi).
\end{equation}

{\bf Proposition 7.3.} {\it Let $\rho$ be an arbitrary
$J$-invariant measure with the zero mean value. Then}
\begin{equation}
\label{GSI}\rm{cov}^c \rho = 2 \rm{cov} \; \rho
\end{equation}

{\bf Proof.} We have
$$
\rm{cov}^c \rho (y, y)= \int_\Om \vert < y, \psi>\vert^2
d\rho(\psi)= \int_\Om \vert (y, \psi) - i w(y, \psi)\vert^2
d\rho(\psi)
$$
$$
= \int_\Om [(y, \psi)^2 + (y, J\psi)^2] d\rho(\psi).
$$
By using symplectic invariance of the measure $\rho$ we get:
$$
\int_\Om (y, J\psi)^2 d\rho(\psi)= \int_\Om (y, \psi)^2 d\rho(\psi).
$$
Thus
$$
\rm{cov}^c \rho (y, y)= 2 \int_\Om (y, \psi)^2 d\rho(\psi)= 2
\rm{cov} \rho (y, y).
$$

\medskip

{\bf Theorem 7.1.} {\it For any measure $\rho$ with the zero mean
value and any $J$-commuting operator $A$, we have:
\begin{equation}
\label{GSIZ} \int_\Omega < A\psi, \psi > d\rho(\psi) = \rm{Tr} \;
\rm{cov}^c \rho \; A;
\end{equation}
in particular, }
\begin{equation}
\label{GSI0}
\sigma^2(\rho) = \rm{Tr}\;  \rm{cov}^c \rho.
\end{equation}

{\bf Proof.} Let $\{e_j\}$ be an orthonormal basis in $\Omega_c$ (we
emphasize that orthogonality and normalization are with respect to
the complex and not real scalar product). Then:
$$
\rm{Tr}\;  \rm{cov}^c \rho\; A =
 \int_\Om \sum_j  < Ae_j, \psi> <\psi, e_j> d\rho(\psi)=
\int_\Omega <A\psi, \psi> d\rho(\psi).
$$

\medskip

We recall that we showed in  [1] that
and
\begin{equation}
\label{GSI1}
\sigma^2(\rho) = \rm{Tr}\; \rm{cov} \rho.
\end{equation}

It seems that there is a contradiction
between equalities (\ref{GSI1}), (\ref{GSI0}) and (\ref{GSI}). In fact, there is no contradiction,
 because in (\ref{GSI1}) and (\ref{GSI0}) we use two different  traces: with respect to the real and complex
  scalar products, respectively. This  is  an important point; even normalization by trace one for the
  von Neumann density operator is the normalization with respect to the complex scalar
  product. By using indexes ${\bf R}$ and ${\bf C}$ to denote real
  and complex traces, respectively, we can write:
  $$
  \sigma^2(\rho)= \rm{Tr}_{{\bf R}} \; \rm{cov}\;  \rho =
  \rm{Tr}_{{\bf C}} \; \rm{cov}^c\;  \rho.
  $$
We remark that  {\it the complex average $m_\rho^c$ and the
covariation operator $B^c$  are ${\bf C}$-linear even if a measure
is not $J$-invariant.} However, in general real and complex averages
do not coincide and real and complex covariance operators are not
coupled by (\ref{GSI}).

Let us find relation between $B={\rm cov} \rho$ and $B^c={\rm cov}^c \rho$ in the general case. It is easy to see that for
\[ B=\left( \begin{array}{ll}
B_{11} & B_{12}\\
B_{21} & B_{22}
\end{array}
\right ), B^*_{11}=B_{11}, B_{22}^*=B_{22}, B_{12}^*=B_{21}
\]
and
\[ B^c=\left( \begin{array}{ll}
D & S\\
-S & D
\end{array}
\right )
\]
we have

{\bf Proposition 7.4.}
{\it The blocks in real and complex covariation operators are connected by the following equalities:}
\begin{equation}
\label{COM}
D=B_{11} + B_{22}, S=B_{12}-B_{21}.
\end{equation}

Thus {\it in the general case the complex covariation operator $B^c$ does not determine the Gaussian measure $\rho_B$ uniquely.}

Let now $\rho_B$ be $J$-invariant. Then
\[ B=\left( \begin{array}{ll}
B_{11} & B_{12}\\
-B_{12} & B_{11}
\end{array}
\right ).
\]
Thus
\begin{equation}
\label{COM1}
D=2 B_{11}, S= 2 B_{12},
\end{equation}
so we obtain (\ref{GSI}) and, hence, we obtain:

\medskip

{\bf Corollary 7.2.} {\it There is one-to-one correspondence between
$J$-invariant Gaussian measures with the zero mean value and complex
covariation operators.}\footnote{These are ${\bf C}$-linear
self-adjoint positively defined operators $B^c: \Omega_c \to
\Omega_c$ belonging to the trace class}

As was remarked, even for a Gaussian measure $\rho$ which is not
$J$-invariant its complex covariation operator $B^c$ does not define
$\rho$ uniquely. Nevertheless, let us represent an arbitrary measure
$\rho$ (with zero mean value and finite dispersion)  by its complex
covariation operator $B^c$ (so we ``project'' measures to their
complex covariation operators).

  Let us consider the dynamics of $\rho$ induced by a dynamics in
  $\Omega$ with a quadratic $J$-invariant Hamilton function ${\cal
  H}.$ We obtain a one-parameter family of measures $\rho_t= U_t^*
  \rho.$ It is easy to see that $B^c(t)= U_t B^c U_t^*.$ Since
  $[B^c, J]=0,$ the $B^c(t)$ satisfies the von Neumann equation
  (\ref{DHMX}).

\section{Prequantum classical statistical model}

We consider the infinite-dimensional phase-space (space of classical
fields) $\Om= Q\times P,$ where $Q$ and $P$ are copies of the
(separable) Hilbert space.  Our aim is to construct a prequantum
classical statistical model on this phase-space inducing the
conventional (Dirac-von Neumann) quantum statistical model
$$
N_{\rm{quant}}=({\cal D}(\Omega_c), {\cal L}_{\rm{s}}(\Omega_c)),
$$
where the complex Hilbert space $\Omega_c= Q \oplus i P.$ Here
${\cal D}(\Omega_c)$ is the space of density operators and ${\cal
L}_{\rm{s}}(\Omega_c)$ is the space of bounded self-adjoint
operators in $\Omega_c$ (quantum observables).\footnote{To simplify
considerations, we consider only quantum observables represented by
bounded operators. To obtain the general quantum model with
observables represented by unbounded operators, we should consider a
prequantum classical statistical model based on the Gelfand triple:
$\Omega_c^{+} \subset \Omega_c \subset \Omega_c^{-}.$}

We choose the space of classical statistical states $S_{G,
\rm{symp}}^\alpha(\Omega)$ consisting of $J$-invariant Gaussian
measures having zero mean value and dispersion $\alpha.$

We choose, cf. [1], the space of classical physical variables as the
functional space ${\cal V}_{\rm{symp}}(\Omega)$ consisting of real
analytic functions, $f:\Omega \to {\bf R},$ that have the
exponential growth:
\begin{equation}
\label{COR1}\mbox{there exist}\;  C_0, C_1 \geq 0 :  \vert
f(\psi)\vert \leq C_0 e^{C_1 \Vert x\Vert};
\end{equation}
preserve the state of vacuum:
\begin{equation}
\label{COR2}
f(0)=0
\end{equation}
and that are $J$-invariant: $f(J\psi)= f(\psi).$

We pay attention that any  $f \in {\cal V}_{\rm{symp}}(\Omega)$ is
an {\it even function:} $f(-\psi)= f(J^2 \psi)= f(J\psi)=f(\psi).$
We shall also use a simple consequence of this result: if $f\in
{\cal V}_{\rm{symp}}(\Omega),$ then its derivative is an odd
function.

{\bf Example 8.1.} Let ${\bf H} \in {\cal L}_{\rm{symp,
s}}(\Omega).$
 Then any polynomial $f(\psi)= \sum_{k=1}^N a_k ({\bf H} \psi,
 \psi)^k, a_k \in {\bf R},$ belongs to the space ${\cal
 V}_{\rm{symp}}(\Omega).$

The following trivial mathematical result plays the fundamental role
in establishing classical $\to$ quantum correspondence.

{\bf Proposition 8.1.} {\it Let $f \in {\cal
V}_{\rm{symp}}(\Omega).$ Then }
\begin{equation}
\label{COR3}  f^{\prime \prime}(0)\in {\cal L}_{\rm{symp},
s}(\Omega).
\end{equation}

We remark that for an arbitrary $\psi \in \Omega$ we have
\begin{equation}
\label{COR3Z} J  f^{\prime \prime}(\psi) =  f^{\prime \prime}(J\psi)
J. \end{equation}

\medskip

We consider now the classical statistical model:
\begin{equation}
\label{CL} M_{\rm{symp}}^\alpha= ( S_{G, \rm{symp}}^\alpha(\Omega),
{\cal V}_{\rm{symp}}(\Omega) ).
\end{equation}

Let us find the average of a variable $f \in {\cal
V}_{\rm{symp}}(\Omega)$ with respect to a statistical state $\rho_B
\in S_{G,\rm{symp}}^\alpha(\Omega):$
$$
<f>_{\rho_B}= \int_\Omega f(\psi) d\rho_B(\psi) = \int_\Omega
f(\sqrt{\alpha} \psi) d\rho_D (\psi)
$$
\begin{equation}
\label{ANN1} =\sum_{n=2}^\infty  \frac{(\alpha)^{n/2}}{n!}
\int_\Omega f^{(n)}(0)(\psi, ...,\psi)d\rho_D (\psi),
\end{equation}
where the covariation operator of the $\sqrt{\alpha}$-scaling
$\rho_D$ of the Gaussian measure $\rho_B$ has the form:
$$
D=B/\alpha.
$$
Since $\rho_B\in S_G^\alpha(\Omega),$ we have $\rm{Tr} \; D  = 1.$
The change of variables in (\ref{ANN1}) can be considered as scaling
of the magnitude of statistical  (Gaussian) fluctuations.
Fluctuations which were considered as very small,
\begin{equation}
\label{DS2} \sigma^2 (\rho)= \alpha,
\end{equation}
(where $\alpha\to 0$ is a small parameter) are considered in the new
scale as standard normal fluctuations.\footnote{Thus QM is a kind of
the statistical microscope which gives us the possibility to see the
effect of fluctuations of  the magnitude $\alpha$ in a neighborhood
of vacuum field point, $\psi_{\rm{vacuum}}\equiv 0)$.} By
(\ref{ANN1}) we have:
\begin{equation}
\label{ANN2} <f>_\rho=  \frac{\alpha}{2}  \int_\Omega (f^{\prime
\prime}(0)\psi, \psi) d\rho_D(\psi) + o(\alpha), \; \alpha \to 0,
\end{equation}
or
\begin{equation}
\label{ANN3}<f>_\rho =   \frac{\alpha}{2} \; \rm{Tr}\; D \;
f^{\prime \prime}(0) + o(\alpha), \; \alpha \to 0.
\end{equation}

Finally, we rewrite the formulas (\ref{ANN2}) and (\ref{ANN3}) in
the complex form:
\begin{equation}
\label{ANN20} <f>_\rho=   \frac{\alpha}{2} \; \int_\Omega <f^{\prime
\prime}(0)\psi, \psi> d\rho_D(\psi) + o(\alpha), \; \alpha \to 0,
\end{equation}
or \begin{equation} \label{ANN30} <f>_\rho =  \alpha  \; \rm{Tr}\;
D^c \;  \frac{f^{\prime \prime}(0)}{2} + o(\alpha), \; \alpha \to 0.
\end{equation}
We pay attention that in (\ref{ANN3}) a trace is the trace with
respect to the real scalar product and in (\ref{ANN30}) - the
complex scalar product.

For a classical variable $f(\psi),$ we define its amplification by
$$
f_\alpha(\psi)= \frac{1}{\alpha} f(\psi)$$ (when $\alpha\to 0$ this
amplification will be becoming infinitely large).
 We see that the classical average of the amplification $f_\alpha(\psi)$ of a classical variable
$f(\psi) $(computed in the model (\ref{CL}) by using the
measure-theoretic approach) is approximately equal to the quantum
average (computed in the model $N_{\rm{quant}}=({\cal D}(\Omega_c),
{\cal L}_{\rm{s}}(\Omega_c))$ with the aid of the von Neumann
trace-formula):
\begin{equation}
\label{ANN30A}<f_\alpha>_\rho =  \rm{Tr}\; D^c \;\frac{f^{\prime
\prime}(0)}{2} + o(1), \; \alpha \to 0.
\end{equation}

The classical $\to$ quantum correspondence map $T$ is similar to the
map presented in [1] in the real case:
\begin{equation}
\label{TT1}T: S_{G, \rm{symp}}^\alpha(\Omega) \to {\cal
D}(\Omega_c), \; \; T(\rho)= \frac{\rm{cov}^c \rho}{\alpha}
\end{equation}
\begin{equation}
\label{TT2}T: {\cal V}_{\rm{symp}}(\Omega) \to {\cal
L}_{\rm{s}}(\Omega_c), \; \; T(f)= \frac{f^{\prime \prime}(0)}{2}
\end{equation}

\medskip

{\bf Theorem 8.2.} {\it The map $T,$ given by (\ref{TT1}),
(\ref{TT2}), establishes a projection of the classical statistical
model $M_{\rm{symp}}^\alpha$ onto the Dirac-von Neumann quantum
model $N_{\rm{quant}}.$ The map (\ref{TT1}) is one-to-one
(bijection); the map (\ref{TT2}) is only onto (surjection). The
latter map is a ${\bf R}$-linear operator. Classical and quantum
averages are coupled via the asymptotic equality (\ref{ANN30}).}

\medskip

We remark that our projection map $T: {\cal
V}_{\rm{symp}}(\Omega)\to {\cal L}_{\rm{s}}(\Omega_c)$ fulfills an
important  postulate for classical$\to$quantum correspondence which
was used by J. von Neumann:
\begin{equation}
\label{TT2Z} T(\sum \lambda_j f_j) = \sum \lambda_j T(f_j),\; \;
\lambda_j \in {\bf R}, f_j \in {\cal V}_{\rm{symp}}(\Omega).
\end{equation}
Here quantum observables $A_j=T(f_j)$ can be incompatible, so these
operators can be noncommuting, see von Neumann [10].
 This postulate was strongly criticized by J. Bell [23] and L.
Ballentine [24] as nonphysical -- because it is not easy to give a
physical meaning to a linear combination of incompatible
observables. I agree that their arguments are not meaningless and
there can be really problems with an experimental realization of the
right-hand side of (\ref{TT2Z}). But in a theoretical model the
relation (\ref{TT2Z}) might be in principle well established.
Therefore, in spite  the critical arguments  of Bell and Ballentine,
there is nothing ``pathological'' in this relation. We recall that
our projection $T$ is not one-to-one on the space of physical
variables, but von Neumann postulated that a such correspondence
should be one-to-one. Nevertheless, we have:

{\bf Corollary 8.1.}  {\it The restriction of the classical $\to$
quantum map $T$ onto the space of quadratic $J$-invariant variables
$V_{\rm{quad, symp}}(\Omega)$ is one-to-one map with its image
${\cal L}_s(\Omega_c).$}

\medskip

{\bf Remark 8.1.} (Quadratic classical variables) Corollary 8.1
shows that the classical $\to$ quantum map $T: V_{\rm{quad,
symp}}(\Omega)\to {\cal L}_s(\Omega_c)$ is nondegenerate. Each
quantum observable $A$ has uniquely defined classical preimage
$f(\psi)= \frac{1}{2} (A\psi, \psi).$ In principle, we could choose
the classical statistical model:
\begin{equation}
\label{CL1} M_{\rm{quad, symp }}^\alpha= ( S_{G,
\rm{symp}}^\alpha(\Omega), V_{\rm{quad, symp}}(\Omega)).
\end{equation}
There is one-to-one correspondence between elements of this
classical model and the Dirac-von Neumann model. Another important
argument to choose this classical model is that the Schr\"odinger
dynamics is in fact dynamics for a quadratic $J$-invariant Hamilton
function. Nevertheless, we do not restrict our consideration to the
classical model $M_{\rm{quad, symp }}^\alpha.$  We can speculate
that linearity of the Hamilton-Schr\"odinger evolution is just an
approximative linearity of nonlinear dynamics in $\Omega$ induced by
nonquadratic Hamilton functions. But this interesting problem should
be investigated in more detail.

{\bf Remark 8.2.} (On the choice of a space of classical variables)
We chosen the functional space ${\cal V}_{\rm{symp}}(\Omega)$ by
generalizing the class of quadratic forms $V_{\rm{quad,
symp}}(\Omega)).$ As the main characteristic for generalization we
chosen the condition of $J$-invariance. By Proposition 8.1 this
condition implies that, for $f \in {\cal V}_{\rm{symp}}(\Omega),$
its second derivative is a $J$-commuting operator. However, such a
choice of the functional space of classical physical variables is
not unique. There can be chosen other characteristics of quadratic
forms $f\in V_{\rm{quad, symp}}(\Omega))$ to obtain  spaces of
classical variables different from ${\cal V}_{\rm{symp}}(\Omega)$
and, nevertheless, reproducing the class of quantum observables. The
problem of an adequate choice of a space of classical variables (as
well as statistical states) is not yet solved, see also section 12
for further considerations.

\section{Gaussian measures inducing quantum pure states}

Let  $\Psi=u + iv \in \Omega_c, $ so $u \in Q, v \in P$ and let
$||\Psi||=1.$ By using the conventional terminology of quantum
mechanics we say that such a normalized vector of the complex
Hilbert space $\Psi$ represents a {\it pure quantum state.} By
Born's interpretation of the wave function a pure state $\Psi$
determines the statistical state with the density matrix:
\begin{equation}
\label{DM}D_\Psi=\Psi \otimes \Psi
\end{equation}
This Born's interpretation of the vector $\Psi$ -- which is, on one
hand, the pure state $\Psi \in \Omega_c$ and, on the other hand, the
statistical state $D_\Psi$ -- was the root of appearance in QM such
a notion as individual (or irreducible) randomness. Such a
randomness could not be reduced to classical ensemble randomness,
see von Neumann [5].

In our approach the density matrix $D_\Psi$ has nothing to do with
the individual state (classical field). The density matrix $D_\Psi$
is the image of the classical statistical state -- the $J$-invariant
Gaussian measure $\rho_\Psi$ on the phase space\footnote{This
measure is uniquely defined, see Proposition 7.3.} having zero mean
value and the (complex) covariation operator
\begin{equation}
\label{BM} B_\Psi^c=  \alpha D_\Psi
\end{equation}
or
\[B_\Psi^c= 2 \alpha \left( \begin{array}{ll}
 u \otimes u + v \otimes v & v \otimes u-u \otimes v\\
 u \otimes v-v \otimes u & u \otimes u + v \otimes v
 \end{array}
 \right ).
 \]
In measure theory there is  used the real covariation operator $B.$
As we know (see Proposition 7.3), for a $J$-invariant Gaussian
measure the real and complex covariation operators are coupled by
the equality:
\begin{equation}
\label{BMR} B_\Psi =\frac{1}{2} B_\Psi^c.
\end{equation}
This operator has {\it two real eigenvectors corresponding to the
same eigenvalue} $\lambda= \alpha/2:$
\[ e_\Psi^{(1)}\equiv \Psi= \left( \begin{array}{ll}
  u\\
  v
  \end{array}
  \right ),  \; \;  e_\Psi^{(2)}\equiv i\Psi= \left( \begin{array}{ll}
      -v\\
      \; \;  u
      \end{array}
      \right ).
        \]

 Thus the Gaussian measure $\rho_\Psi$ has the support in the real plane
$$
\Pi_\Psi=\{\psi=x_1 e_\Psi^{(1)} + x_2 e_\Psi^{(2)}: x_j \in \br\}
$$
and
$$
d \rho_\Psi (x_1, x_2)=\frac{1}{\pi \alpha} e^{- \frac{x_1^2 +
x_2^2}{\alpha}} dx_1 dx_2.
$$
We remark that for two dimensional Gaussian distributions symplectic
invariance is equivalent to coincidence of eigenvalues of the
covariance matrix, i.e., $(p \leftrightarrow q) -$ symmetry of
Gaussian distribution: $B=\rm{diag} (\alpha/2, \alpha/2).$

\medskip

{\bf Physical consequence.} {\it There are no ``pure quantum
states.'' States that are interpreted in the conventional quantum
formalism as pure states, in fact, represent $J$-invariant Gaussian
measures having two dimensional supports. Such states can be
imagined as fluctuations of fields concentrated on two dimensional
real planes of the infinite dimensional state phase-space.}

\medskip

We recall that in quantum theory one distinguishes so called {\it
pure states} and so called {\it mixtures.} Let us discuss this point
in more detail. The set of density operators ${\cal D}(\Omega_c)$ is
a positive cone in the space of all trace class operators: if $D_1,
D_2 \in {\cal D}(\Omega_c), $ then $p_1 D_1 + p_2 D_2 \in {\cal
D}(\Omega_c)$ for any $p_1, p_2 \geq 0, p_1 + p_2=1.$ We recall, see
e.g. [6], that the set of extreme points of the cone ${\cal
D}(\Omega_c)$ coincides with the set of pure states. Thus only pure
states $D_\psi$ could not be represented in the form of a
statistical mixture:
$$
D=p_1 D_1 + p_2 D_2, p_j > 0, p_1 + p_2=1.
$$
It seems that this mathematical result was one of the reasons why J.
von Neumann distinguished sharply pure states and statistical
mixtures and elaborated the notion of {\it individual randomness} -
randomness associated with ``pure states,'' see [25]. In our
approach there is no difference between ``pure quantum states'' and
``quantum statistical mixtures'' (at least from probabilistic
viewpoint; geometry of distributions corresponding to ``pure
states'' is very special; they are concentrated on two dimensional
real subspaces).
\medskip

{\bf Example 9.1.} Let us consider the classical statistical state
(Gaussian measure) $\rho \equiv \rho_\Psi$ which is projected onto
the ``pure quantum state'' $\Psi \in \Omega, ||\Psi ||=1.$ The
measure $\rho$ is concentrated on the real plane $\Pi_\Psi.$ Thus we
can restrict our considerations to the phase space $\Omega=\br
\times \br$ and the measure
$$
d \rho(q, p)=\frac{1}{\pi \alpha} e^{-\frac{1}{ \alpha}(p^2 + q^2)}.
$$
Let us consider a $J$-invariant physical variable $$ f(q,
p)=\frac{1}{2} [(p^2 + q^2) + (p^2 + q^2)^2].
$$
We have
$$
<f>_\rho= \frac{1}{2} \int \int (p^2 + q^2) d \rho (q, p) +
\frac{1}{2} \int \int (p^2 + q^2)^2 d \rho(q, p)= \alpha I_3 +
\alpha^2 I_5,
$$
where $I_n= 2\int_0^\infty s^n e^{-s^2} ds.$ Now we make the
amplification of the classical variable $ f_\alpha(q,
p)=\frac{1}{2\alpha} [(p^2 + q^2) + (p^2 + q^2)^2]$ and obtain:
$$
<f_\alpha>_\rho=  I_3 + \alpha I_5.
$$
Thus approximately $<f_\alpha>_\rho$ is equal to $I_3$ - the quantum
average. This is the essence of quantum averaging: only the
quadratic part $f_{\rm{quad}}(q,p)= \frac{1}{2} (p^2 + q^2)$ of a
physical variable $f(q,p)$ is taken into account; the contribution
of terms of higher orders is neglected.

\section{Prequantum classical statistical field theory (PCSFT)}

Let $Q=P=L_2 ({\bf R}^3)$ be the Hilbert space of real valued square
integrable functions $\psi:{\bf R}^3 \to {\bf R}$ with the scalar
product $(\psi_1, \psi_2)=\int_{{\bf R}^3} \psi_1(x) \psi_2 (x) dx.$
 Our classical phase space
$\Omega=L_2({\bf R}^3) \times L_2({\bf R}^3)$ consists of vector
functions $\psi(x)=\left(
\begin{array}{ll}
q(x)\\
p(x)
\end{array}
\right ). $  The symplectic operator $J$ on this phase-space  has
the form:
\begin{equation}
\label{EQQ} q_1(x)= p(x), \; p_1(x)= -q(x)
\end{equation}
and the symplectic form on $\Omega$ is defined by $w(\psi_1,
\psi_2)=\int_{{\bf R}^3} (p_2 (x) q_1(x) - p_1(x) q_2 (x)) dx.$ The
fundamental law of PCSFT is the invariance of physical variables
with respect to this transformation. By introducing on $\Omega$ the
canonical complex structure we obtain the  $\Omega_c =L_2^{\bf C}
({\bf R}^3)$ -- the complex Hilbert space of square integrable
functions $\psi:{\bf R}^3 \to {\bf C}, \psi=q(x) + ip(x))$ with the
scalar product $(\psi_1, \psi_2)=\int_{\bf R}^3
\psi_1(x)\bar{\psi}_2 (x) dx.$ Let us consider an integral operator
$$
A:\Omega \to \Omega, A \psi(x)=\int_{{\bf R}^3}A(x, y) \psi(y) dy.
$$
The kernel $A(x, y)$ of such an operator has the block structure.
This operator is $J$-invariant iff $A_{11}(x, y)=A_{22}(x, y),
A_{12}(x, y)=-A_{21}(x, y),$ and it is symmetric iff $A_{11}(x,
y)=A_{11}(y, x), A_{12}(y, x)=A_{21}(x, y)=-A_{12}(x, y).$ The
corresponding quadratic form
$$
f(\psi)=\frac{1}{2}\Big[\int A_{11} (x, y) \psi_1 (x) \psi_2 (y) dx
dy + 2 \int A_{12} (x, y) \psi_2 (x) \psi_1 (y)
$$
$$
+ \int A_{11} (x, y) \psi_2 (x) \psi_2 (y) dx dy\Big]
$$
Let $\rho$ be a $J$-invariant measure on $\Omega= L_2({\bf R}^3)
\times L_2({\bf R}^3).$ Its complex covariance is defined by:
$$
<B^c \psi_1, \psi_2>=\int_{L_2({\bf R}^3) \times L_2({\bf R}^3)}
\Big( \int_{{\bf R}^3} \psi_1 (x) \bar{\psi}(x) dx \int_{{\bf
R}^3}\psi (x) \bar{\psi}_2 (x) dx\Big) d\rho (\psi).
$$
Let $\rho$ has the dispersion $\sigma^2 (\rho)=\int_ {L_2({\bf R}^3)
\times L_2({\bf R}^3)} (\int_{{\bf R}^3}|\psi(x)|^2 dx) d\rho
(\psi)=\alpha.$  We find the average of the quadratic physical
variable $f$ in the state $\rho:$
$$
<f>_\rho= \frac{1}{2}  \int_{L_2({\bf R}^3) \times L_2({\bf R}^3)}
\Big( \int A_{11} (x, y) q(x) q(y) dx dy
$$
$$ + 2 \int A_{12} (x,
y) p (x) q(y) dx dy + \int A_{11} (x, y) p(x) p(y) dx dy\Big)
d\rho_B (q, p) $$
$$
= {\rm Tr} B^c A= \alpha {\rm Tr} D^c\;
\frac{f^{\prime\prime}(0)}{2},
$$
where $D^c= B^c/ \alpha$ is the von Neumann density operator
obtained through the scaling of the  covariation operator of the
Gaussian measure $\rho$ representing a classical statistical state.
 Since
the observable is quadratic, there is the precise equality of the
average of the $\frac{1}{\alpha}$-amplification of the classical
variable $f$ and the quantum average of the self-adjoint operator
$2A= f^{\prime\prime}(0)/2.$

 Let us forget for a moment about mathematical difficulties
and consider a singular integral operator - differential operator:
$$
{\bf H}=- \frac{ \Delta}{2m} + V(x)
$$
We consider in phase-space $\Omega$ the diagonal operator ${\bf
H}_{11}={\bf H}_{22}={\bf H}, {\bf H}_{12}={\bf H}_{21}=0$ or we can
directly consider ${\bf H}$ as acting in the complex Hilbert space
$\Omega_c=L_2^{\bf C} ({\bf R}^3).$ The corresponding classical
Hamilton function is quadratic:
$$
{\cal H}(\psi)=\frac{1}{2} <{\bf H}\psi, \psi>= \frac{1}{2}
\int_{{\bf R}^3}\Big(\frac{|\nabla \psi (x)|^2}{2m}  + V(x) |\psi
(x)|^2\Big) dx.
$$
This is the classical energy of a system with the infinite-number of
degrees of freedom -- the system of coupled fields $q(x)$ and
$p(x).$ This system is a classical vector field; the parameter $m$
-- ``mass'' -- is one of characteristics of this field. ${\cal
H}(\psi)$ is an ordinary function (functional) of $\psi.$ We can
find its classical average:
$$<{\cal H}>_\rho
=\frac{1}{2}\int_{L_2({\bf R}^3) \times L_2({\bf R}^3)}<{\bf H}\psi,
\psi> d\rho (\psi) $$ $$ = \frac{1}{2} \int_{L_2({\bf R}^3) \times
L_2({\bf R}^3)} \Big(\frac{1}{2} \int_{{\bf R}^3}(\frac{|\nabla \psi
(x)|^2}{2m}
 + V(x) |\psi (x)|^2) dx\Big)d\rho_B (\psi)
$$
$$
= \frac{1}{2}\rm{Tr}  \; B^c  \; \Big(\frac{-\Delta}{2m}  +
V(x)\Big)= \alpha \rm{Tr} \; D^c {\bf H}, \; \; \mbox{where} \; \;
D^c=B^c/\alpha.
$$
Of course, we understood that, since the operator ${\bf H}$ is
unbounded, the $\rm{Tr} \; D^c {\bf H}$ is not well defined for an
arbitrary Gaussian measure. One of possible solutions of this
problem is to choose the class of Gaussian measures depending on the
quantum operator. Another possibility is to follow J. von Neumann
[10] and consider an approximation of ${\bf H}$ by bounded operators
representing {\it unsharp measurement} of energy.

We emphasize again that we could not guarantee that the quantum
observable of energy ${\bf H}$ really corresponds to a quadratic
classical variable of energy ${\cal H}(\psi)= <{\bf H}\psi, \psi>$
(in fact, to its amplification ${\cal H}_\alpha(\psi)=\frac{1}{2
\alpha}<{\bf H}\psi, \psi>).$ Let us e.g. the classical
energy-variable of the form:
$$
{\cal F}(\psi)= \Big[\frac{1}{2} \int_{{\bf R}^3}(\frac{|\nabla \psi
(x)|^2}{2m}   + V(x) |\psi (x)|^2)+ g \int_{{\bf R}^3} |\psi (x)|^4
dx\Big], \; g>0
$$
(which is $J$-invariant). Then it produces the same quantum average
as the quadratic energy-variable:
$$
<{\cal F}>_\rho = \alpha \rm{Tr} \; D^c {\bf H} + \alpha^2 g
\int_{L_2({\bf R}^3) \times L_2({\bf R}^3)} \Big(\int_{{\bf R}^3}
|\psi (x)|^4 dx\Big) d\rho_D(\psi)
$$
 Of course, the
latter Hamilton function will induce nonlinear Hamiltonian dynamics
in the infinite-dimensional phase-space $\Omega,$ and in principle
it could be distinguished from the  linear dynamics. We now consider
the quadratic classical variables inducing quantum observables of
the position ${\bf x}_j$  and the momentum ${\bf p}_j (j=1,2,3):$
$$
f_{{\bf x}_j}(\psi)=\frac{1}{2} <{\bf x}_j \psi, \psi>= \frac{1}{2}
\int_{{\bf R}^3} x_j |\psi (x)|^2 d x.
$$
$$
f_{{\bf p}_j}(\psi)=\frac{1}{2} <{\bf p}_j \psi, \psi>= \frac{1}{2}
\int_{{\bf R}^3} y_j |\tilde{\psi}(y)|^2 d y,
$$
where $\tilde{\psi}(y)$ is the Fourier transform of the
$L_2$-function $\psi(x).$ We can also consider the quadratic
classical variables inducing the angular momentum operators, e.g.,
$$
f_{{\bf J}_z}(\psi)=\frac{1}{2} <{\bf J}_z \psi, \psi>= \frac{-i}{2}
\int_{{\bf R}^3}\Big( x\frac{\partial \psi}{\partial y} - y
\frac{\partial \psi}{\partial x}\Big) \bar{\psi} \; d x dy d z
$$
(but the same quantum operator  also can be induced e.g. by the
classical variable: $f_{{\bf J}_z}(\psi)=\frac{1}{2} \Big(<{\bf J}_z
\psi, \psi> + <{\bf J}_z \psi, \psi>^2\Big).$

\section{Fundamental field}
In section 10 we proposed the interpretation of PCSFT by which,
instead of particles, we considered corresponding fields, e.g., the
electron field. Each field $\psi(x)=(q(x), p(x))$ evolves as a pair
of self-inducing fields and the system of Hamiltonian equations
(\ref{HF1}) describes its motion. We consider the nonrelativistic
case and scalar fields $q(x)$ and $p(x).$ In this case the Hamilton
function has the form (\ref{HF3}): ${\cal H}(\psi)=\int_{\br^6} R(x,
y) \psi(x) \bar{\psi}(y) dx dy,$ where
\begin{equation}
\label{PS} R(x, y)=-\frac{\nabla^2 \delta (x-y)}{2m}  + \delta (x-y)
V(x) .
\end{equation}

In section 10 we interpreted $m$ as a parameter, mass, determining a
scalar-complex field (or a pair of self-inducing real fields); the
potential $V(x, y)=\delta (x-y) V(x)$ was considered as an external
potential contributing into a self-interaction of
$\psi(x).$\footnote{ In fact, the component $q(x)$ self-interact
with itself; the same is valid for $p(x);$ Here are no
cross-interactions. This self-interaction is local, since it
contains the $\delta$-function.}

Such an interpretation of PCSFT was based on splitting of the kernel
$R(x, y)$ into two summands, see (\ref{PS}), and on different
interpretation of these summands. The first was considered as an
internal contribution of the field and the second as an an external
potential.

We now propose a new interpretation. We consider also the first
summand in (\ref{PS})  as an external potential inducing a
self-interaction of the field $\psi(x).$

{\bf Definition 11.1.} {\it A mass interaction - field
(corresponding to the mass parameter $m > 0$) is defined as}
\begin{equation}
\label{PS1} R_m (x, y)=-\frac{\nabla^2 \delta (x-y)}{2m}
\end{equation}

\medskip

\centerline{{\bf Fundamental Field Interpretation:}}

\medskip

There is the unique {\it fundamental} field $\psi(x)=(q(x), p(x))$
which interact with various potentials.\footnote{In our mathematical
model an interaction potential $R(x, y)$ can be any distribution on
$\br^6.$ } In the conventional model an interaction potential $R(x,
y)$ is always of the form (\ref{HF3}). So it contains a mass
interaction field.\footnote{ We recall that we consider
nonrelativistic fields, so $m > 0.$} Thus we propose to the
following interpretation of PCSFT:

a). There is the fundamental vector-field $\psi(x)=(p(x), q(x)).$

b). Its internal (ontic) energy is given by:
\begin{equation}
\label{E} {\cal H} (\psi)=\frac{1}{2} ||\psi||^2=\frac{1}{2}
\int_{\br^3} (q^2 (x) + p^2 (x)) dx.
\end{equation}

c). There are various interaction-fields $R(x, y)$ inducing
self-interactions of the fundamental field $\psi(x).$

d). The energy of the $R$-self-interacting field $\psi(x)$ is given
by:
\begin{equation}
\label{E1} {\cal H}_R(\psi)=\frac{1}{2} \int_{\br^6} R(x, y) \psi
(x) \bar{\psi} (y) dx dy.
\end{equation}

For scalar fields $q(x)$ and $p(x)$  (in the nonrelativistic case)
an interaction field $R(x, y)$ can always be represented in the form
(\ref{PS}), where the first summand is referred to as the mass
interaction-field.

e). In the absence of interaction-fields the fundamental field
$\psi(x)$ evolves as a system with the Hamilton function (\ref{E}):
\begin{equation}
\label{E2} \dot q=p, \dot p=-q
\end{equation}
These are oscillation of the form: $\psi(t, x)=e^{-it} \psi_0
(x).$\footnote{We remark that by the conventional interpretation of
QM functions $\psi(t,x)$ for all $t$  are just representations of
the same pure state that is defined up to $\lambda=e^{i \alpha}.$
But by our interpretation $\psi(t_1,x)$ and $\psi(t_2,x)$ for
$t_1\not= t_2$ are different classical fields.}

e1). In the presence of an interaction-field $R(x, y)$ the
fundamental field $\psi(x)$ evolves as a system with the Hamilton
function (\ref{E1}):
\begin{equation}
\label{E3} \dot q= Rp, \dot p=-Rq,
\end{equation}
where $R \psi(x)=\int_{R^3} R(x, y) \psi(y) dy.$

{\bf Remark 11.1.} By Proposition 5.1 the quadratic form (\ref{E})
of the fundamental field $\psi(x)$ is not changed in the process of
the Hamiltonian evolution for any interaction-field $R(x, y).$ We
call (\ref{E}) {\it internal energy} of $\psi(x).$ In fact, we never
measure the internal energy of the fundamental field $\psi(x).$ We
always measure the energy of $\psi(x)$ corresponding to some
interaction field $R(x, y).$

The main difference between the fundamental $\psi$-field and
interaction $R$-fields is  that for the $\psi$-field we are not able
to prepare individual states (only Gaussian distributions), but for
$R$-fields it is possible to prepare an individual state that can be
unchanged during sufficiently large interval of time. For example,
we are able to prepare the Kulon potential $V(r)=\frac{c}{r}$ and
not only a Gaussian ensemble of such potentials $V(r,
\psi)=\frac{c(\psi)}{r},$ where $\psi$ is a chance parameter. The
same can be said about the mass field. We are able to prepare the
mass potential $R_m(x, y)=\frac{\nabla^2 \delta (x-y)}{2m}$ and not
only a Gaussian ensemble of such potentials $R_m(x,
y)=\frac{\nabla^2 \delta (x-y)}{2m(\psi)}$ (``we are able to create
a particle of the fixed maps $m$'').

For the $\psi$-field we are not able to prepare the fixed state
$\psi_0 (x).$ Even when in quantum mechanics one says that ``a
system is in a stationary (pure) state $\Psi_0$'', in PCSFT this
means just the creation of a Gaussian ensemble of $\Psi$-fields
concentrated on the real plane $\Pi_{\Psi_0}=\{e_1=\Psi_0,
e_2=i\Psi_0\}.$

Finally, we remark that in PCSFT there is  no  difference (from the
physical viewpoint) between the mass potential $R_m (x, y)$ and an
external potential $V(x, y).$ \footnote{One of the purely
mathematical differences is that the mass-potential $R_m(x, y)=
\delta (x- y) \delta^{\prime\prime} (x)$ is more singular compared
to $R_V (x, y)=\delta(x-y) V(x),$ where $V(x)$ is typically a piece
wise smooth function. But, of course, there can be considered
singular potentials $V(x)$, e.g., $V(x)=\delta(x).$}

We finish this section with a citation from the book of Einstein and
Infeld [19], p. 242-243: {\small `` But the division into matter and
field is, after the recognition of the equivalence of mass and
energy, something artificial and not clearly defined. Could we not
reject the concept of matter and build a pure field physics? ...
There would be no place in our new physics, for both field and
matter, field being the only reality. This new view is suggested by
the great achievements of field physics, by our success in
expressing the laws of electricity, magnetism, gravitation in the
form of structure laws, and finally by the equivalence of mass and
energy.''}

\section{Dispersion preserving dynamics with nonquadratic Hamilton functions}

By considering nonquadratic observables, see section 8, we come to a
new interesting problem: investigation of dynamics with nonquadratic
Hamilton functions. Let us consider an arbitrary Hamilton function
${\cal H}: \Om \to \br.$ The first important remark is that such a
dynamics would transfer Gaussian states into Gaussian iff ${\cal H}$
is quadratic.

Suppose that, for any $\psi \in \Om,$ the system of Hamiltonian
equations (\ref{HE}) has the unique
 solution, $\psi(t)\equiv U_t \psi, \psi(0)=\psi.$ In this case there is  well defined the map (Hamiltonian flow)
\begin{equation}
\label{M}
U_t: \Om \to \Om.
\end{equation}
This map induces the map $U_t^*$ in the space of probability
measures $PM(\Omega)$ on the phase-space $\Om,$ see section 4. As
was already mentioned, in the nonquadratic case the measure $U_t^*
\rho, t > 0,$ can be non-Gaussian even for a Gaussian measure
$\rho.$ For nonquadratic Hamilton functions we cannot restrict the
classical statistical model to the model with Gaussian states. We
should consider the space of statistical states consisting of all
probability measures $\rho$ on $\Om$ that have the zero mean value
and the dispersion $\alpha.$ Denote this class by the symbol
$PM^\alpha(\Omega).$ May be we should consider the subclass
$PM^\alpha_{\rm{symp}}(\Omega)$ of  $PM^\alpha(\Omega)$ consisting
of $J$-invariant measures: $J^* \rho= \rho.$ But at the moment we
consider arbitrary measures.

We are interested in Hamiltonian dynamics $U_t$ in the phase space
$\Om$ that induces dynamics $U_t^*$ in $PM^\alpha(\Omega).$ Such a
dynamics preserves the zero mean value and the dispersion $\alpha.$

Quantum dynamics corresponding to the classical Hamiltonian dynamics
with a quadratic $J$-invariant Hamilton functions is an example of
dynamics preserving the zero mean value and the dispersion. We are
interested in more general dynamics with similar features.

We find the dispersion of $U_t^*\rho$ for an arbitrary $\rho\in
PM(\Omega)$ having zero mean value:
\begin{equation}
\label{PR} \sigma^2 (U_t^* \rho)=\int_\Om ||U_t \psi||^2
d\rho(\psi).
\end{equation}
We are interested in a Hamiltonian dynamics such that dispersions of
probability measures are preserved -- {\it dispersion preserving
dynamics.}

Suppose that $U_t$ preserves the mean value of a measure. By
(\ref{PR}) if $U_t$ preserves the norm on the phase space $\Omega$
then $U_t^*$ preserves the dispersion. We remark that a nonlinear
norm preserving map $U:\Om \to \Om$ need not be one-to-one or onto.
Moreover, it need not be an isometry: $\Vert U \psi \Vert = \Vert
\psi\Vert$ for any $\psi \in \Omega$ does not imply that $\Vert U
\psi_1 - U \psi_2 \Vert= \Vert \psi_1 - \psi_2\Vert.$

It is easy to find the sufficient and necessary condition for
norm-preserving dynamics induced by  a Hamilton function ${\cal
H}(\psi).$
 We can write the general Hamiltonian equation (\ref{HE}) in the form:
\begin{equation}
\label{HEF} \dot \psi = J {\cal H}^\prime(\psi).
\end{equation}

{\bf Theorem 12.1.} {\it Let the flow $U_t$ induced by a Hamilton
function ${\cal H}(\psi)$ be a surjection, i.e., $U_t (\Omega)=
\Omega.$ Then it is norm preserving iff the following equality, cf.
(\ref{PWNS1}), section 5, holds}
\begin{equation}
\label{NP} (J {\cal H}^\prime(\psi), \psi)=0, \psi \in \Om.
\end{equation}

{\bf Proof.} a) Let $||U_t \psi||^2=||\psi||^2$ for any $\psi \in
\Om.$ By using the representation (\ref{HEF}) we obtain:
$$
0=\frac{d}{dt}||U_t \psi||^2=2 (J {\cal H}^\prime(U_t \psi), U_t
\psi).
$$
Thus $(J {\cal H}^\prime(U_t \psi), U_t \psi)=0, \psi \in \Om.$ Now
we use the fact that $U_t(\Om)=\Om$ and obtain the equality
(\ref{NP}).

b) Let the equality (\ref{NP}) hold for any point of $\psi\in
\Omega.$ Then, in particular,
\begin{equation}
\label{CONF} (J {\cal H}^\prime(U_t \psi), U_t \psi)=0
\end{equation}
for any $\psi \in \Om.$ Thus $\frac{d}{dt}||U_t \psi||^2=0$ and
hence $||U_t \psi||=||\psi||, t\geq t_0, \psi \in \Om.$

\medskip

We remark that (\ref{NP}) implies norm preserving even in the case
when $U_t$ is not surjection.

Denote the class of maps $f: \Omega \to {\bf R}$ satisfying the
condition (\ref{NP}) by the symbol $W(\Omega).$

{\bf Corollary 12.1.} {\it A Hamiltonian flow is norm preserving iff
the equality (\ref{CONF}) holds.}

\medskip

The equation (\ref {NP}) is a linear equation with respect to ${\cal H}:$
\begin{equation}
\label{NPE} \Big(\frac{\partial {\cal H}}{\partial q},
p\Big)=\Big(\frac{\partial {\cal H}}{\partial p}, q\Big)
\end{equation}

\medskip

{\bf Theorem 12.2.} {\it Let the condition ${\cal H}\in
W(\Omega).$Then ${\cal H}^{\prime\prime}(0) \in {\cal L}_{{\rm
symp},s}(\Omega).$}

{\bf Proof.} We have: $({\cal H}^\prime(\psi), J \psi)=0.$ Thus
${\cal H}^{\prime\prime}(\psi) J \psi + J^* {\cal H}^\prime(\psi)=0$
and, hence, ${\cal H}^{\prime\prime\prime}(\psi)J \psi + {\cal
H}^{\prime\prime}(\psi) J + J^* {\cal H}^{\prime\prime}(\psi)=0.$
Therefore
\begin{equation}
[\label{NPE1} {\cal H}^{\prime\prime}(0), J] =0.
\end{equation}

\medskip
We pay attention that in general we have:
\begin{equation}
[\label{NPE2} {\cal H}^{\prime\prime}(\psi), J] = - {\cal
H}^{\prime\prime\prime}(\psi)J \psi .
\end{equation}

We pay attention that, for any map ${\cal H}: \Omega \to {\bf R},$  we can represent
\[{\cal H}^{\prime\prime}= \left( \begin{array}{ll}
\frac{\partial^2 {\cal H}}{\partial q^2} & \frac{\partial^2 {\cal H}}{\partial q \partial p}\\
\frac{\partial^2 {\cal H}}{\partial p \partial q} & \frac{\partial^2 {\cal H}}{\partial p^2}\\
\end{array}
\right )
\]
 The condition ${\cal H}^{\prime\prime}(0,0) \in {\cal L}_{{\rm symp},s} (\Omega)$  implies that
\begin{equation}
\label{NPEA} \frac{\partial^2 {\cal H}}{\partial
q^2}(0,0)=\frac{\partial^2 {\cal H}}{\partial p^2} (0,0), \;
\frac{\partial^2 {\cal H}}{\partial q \partial p}= -
\frac{\partial^2 {\cal H}}{\partial p \partial q}.
\end{equation}
The latter equality should not be surprising even in the light of the well known equality of mixed partial derivatives
for any two times continuously differentiable map.
Of course, we always have:
$$
\frac{\partial^2 {\cal H}}{\partial p_i \partial q_j}= \frac{\partial^2 {\cal H}}{\partial q_j \partial p_i}
$$
for any $i, j.$ Let us consider an illustrative example. Let us consider the quadratic Hamilton function:
${\cal H} (q_1, q_2, p_1, p_2)=p_1 q_2-q_1 p_2.$ Here we have:
\[\frac{\partial^2 {\cal H}}{\partial q \partial p}= \left( \begin{array}{ll}
\frac{\partial^2 {\cal H}}{\partial q_1 \partial p_1} & \frac{\partial^2 {\cal H}}{\partial q_1 \partial p_2} \\
\frac{\partial^2 {\cal H}}{\partial q_2 \partial p_1} & \frac{\partial^2 {\cal H}}{\partial q_2 \partial p_2} \\
\end{array}
\right )=
\left( \begin{array}{ll}
0 & -1\\
1 & 0
\end{array}
\right );
\]
and
\[\frac{\partial^2 {\cal H}}{\partial p \partial q}= \left( \begin{array}{ll}
\frac{\partial^2 {\cal H}}{\partial p_1 \partial q_1} & \frac{\partial^2 {\cal H}}{\partial p_1 \partial q_2} \\
\frac{\partial^2 {\cal H}}{\partial p_2 \partial q_1} & \frac{\partial^2 {\cal H}}{\partial p_2 \partial q_2} \\
\end{array}
\right )=
\left( \begin{array}{ll}
0 & 1\\
-1 & 0
\end{array}
\right )
\]

We remark that any polynomial of type considered in Example 8.1
satisfies the condition (\ref{NP}). Therefore each Hamilton function
of such a type induces the flow $U_t(\psi)$ that preserves the norm,
e.g., ${\cal H}(\psi)= a_1({\bf H} \psi, \psi) + a_2 ({\bf H} \psi,
\psi)^2,$ where $[{\bf H}, J]=0.$ But  we do not know general
relation between the functional classes of $J$-invariant functions
and functions satisfying (\ref{NP}).

On the other hand, by using the condition (\ref{NP}) we can easily
find Hamilton functions that induce flows which do not preserve the
norm. Let us consider (in the two dimensional case) the map ${\cal
H}(q,p) = q^2 p.$ For this map the condition (\ref{NP}) does not
hold true. Therefore the Hamiltonian flow corresponding to this map
does not preserve the norm.

We now investigate conditions for preserving of the average. There
is given a measure $\rho$ with zero mean value (``fluctuation of
vacuum''): $m_\rho=0.$ We would like to find a sufficient condition
for preserving of this value:  $m_{\rho_t} =0$ for $t \geq 0.$ Let
us consider the class of symmetric measures: such $\rho$ that $
g_{-1}^* \rho= \rho,$ where $g_{-1}\psi = -\psi.$ We remark that any
even measure has the zero mean value.

{\bf Proposition 12.1.} {\it Let $\rho$ be a symmetric measure and
let $U_t(\psi)$ be an odd Hamiltonian flow:
\begin{equation}
\label{Z1} U_t(-\psi)=- U_t(\psi).
\end{equation}
 Then the average $m_{\rho_t} =0$ for $t \geq 0.$}

We even can prove that:

{\bf Proposition 12.2.} {\it An odd Hamiltonian flow preserves the
class of symmetric measures.}

{\bf Proof.} We should get $g_{-1}^* U_t^* \rho=U_t^* \rho.$  We
have: $$\int f(\psi) d g_{-1}^* U_t^* \rho (\psi)= \int
f(U_t(-\psi)) d \rho (\psi)= \int f(-U_t(\psi)) d \rho (\psi) $$ $$
= \int f(U_t(\psi)) d \rho (\psi)= \int f(\psi) d U_t^* \rho
(\psi).$$

{\bf Proposition 12.3.} {\it Let the Cauchy problem for a
Hamiltonian equations be well possed. Then the Hamiltonian flow is
odd if ${\cal H}^\prime$ is odd.}

{\bf Proof.} a). Let (\ref{Z1}) hold. Then $\frac{d U_t}{d t}(\psi)=
- \frac{d U_t}{d t}(-\psi).$ Thus ${\cal H}^\prime(U_t(\psi))= -
{\cal H}^\prime(U_t(-\psi)).$ Hence
\begin{equation}
\label{Z2}{\cal H}^\prime(\phi)= - {\cal H}^\prime(-\phi))
\end{equation}
for any $\phi=U_t\psi.$ Since the problem is well possed, any $\phi
\in \Omega$ can be represented in this form.

b). Let now (\ref{Z2}) hold. We have: $-\frac{d U_t}{d t}(-\psi)= -J
{\cal H}^\prime(U_t(-\psi))=  J {\cal H}^\prime(-U_t(-\psi)).$ But
the problem is well possed, so the solution is unique. Thus
(\ref{Z1}) holds.

{\bf Corollary 12.2.} {\it Let the Hamilton function ${\cal
H}(\psi)$ be $J$-invariant. Then its flow preserves the averages of
symmetric measures.}

Finally, we pay attention that any $J$-invariant measure is
symmetric (and in particular its average is zero).

{\bf Corollary 12.3.} {\it Let the Hamilton function ${\cal
H}(\psi)$ and the measure $\rho$ be $J$-invariant. Then the
Hamiltonian flow preserves the (zero) mean value of $\rho.$}

\newpage

{\bf References}

[1] A. Yu. Khrennikov, A pre-quantum classical statistical model
with infinite-dimensional phase space. {\it J. Phys. A: Math. Gen.},
{\bf 38}, 9051-9073 (2005).

[2] L. de la Pena and A. M. Cetto, {\it The Quantum Dice: An
Introduction to Stochastic Electrodynamics Kluwer.} Dordrecht, 1996;
T. H. Boyer, {\it A Brief Survey of Stochastic Electrodynamics} in
Foundations of Radiation Theory and Quantum Electrodynamics, edited
by A. O. Barut, Plenum, New York, 1980; T. H. Boyer, Timothy H.,
{\it Scientific American},pp 70-78, Aug 1985; see also an extended
discussion on vacuum fluctuations in: M. O. Scully, M. S. Zubairy,
{\it Quantum optics,} Cambridge University Press, Cambridge, 1997;
W. H. Louisell, {\it Quantum Statistical Properties of Radiation.}
J. Wiley, New York, 1973; L. Mandel and E. Wolf, {\it Optical
Coherence and Quantum Optics.} Cambridge University Press,
Cambridge, 1995.

[3] L. De La Pena, {\it Found. Phys.} {\bf 12}, 1017 (1982); {\it J.
Math. Phys.} {\bf 10}, 1620 (1969); L. De La Pena, A. M. Cetto, {\it
Phys. Rev. D} {\bf 3}, 795 (1971).

[4] E. Nelson, {\it Quantum fluctuation,} Princeton Univ. Press,
Princeton, 1985.

[5]  G. `t Hooft, ``Quantum Mechanics and Determinism,''
hep-th/0105105.

[6] G. `t Hooft,``Determinism beneath Quantum Mechanics,''
quant-ph/0212095.

[7] A. Yu. Khrennikov, (editor),  {\it Quantum Theory:
Reconsideration of Foundations}-3, American Institute of Physics,
Conference proceedings, {\bf 810}, Melville, New York, 2006.

[8] E. Schr\"odinger,  {\it Philosophy and the Birth of Quantum
Mechanics.} Edited by M. Bitbol, O. Darrigol (Editions Frontieres,
Gif-sur-Yvette, 1992); especially the paper of S. D'Agostino,
``Continuity and completeness in physical theory: Schr\"odinger's
return to the wave interpretation of quantum mechanics in the
1950's'', pp. 339-360.

[9]   E. Schr\"odinger, {\it E. Schr\"odinger Gesammelte
Abhandlungen} ( Wieweg and Son, Wien, 1984); especially the paper
``What is an elementary particle?'', pp. 456-463.

[10]  J. von Neumann,  {\it  Mathematical foundations of quantum
mechanics.} Princeton Univ. Press: Princeton, N.J. (1955).

[11] D. Bohm,  {\it Quantum theory}, Englewood Cliffs, New-Jersey:
Prentice-Hall, 1951.

[12] P. Holland, {\it The quantum theory of motion},  Cambridge:
Cambridge University press,  1993.

[13] A. Yu. Khrennikov, {\it J. Phys.A: Math. Gen.} {\bf 34},
9965-9981 (2001); {\it Il Nuovo Cimento} {\bf B 117},  267-281
(2002); {\it J. Math. Phys.} {\bf 43}, 789-802 (2002);

[14] A. Yu. Khrennikov, {\it Information dynamics in cognitive,
psychological and anomalous phenomena,} Ser. Fundamental Theories of
Physics, Kluwer, Dordreht, 2004.

[15] A. Yu. Khrennikov, {\it J. Math. Phys.} {\bf 44}, 2471- 2478
(2003).

[16]  A. Yu. Khrennikov, {\it Il Nuovo Cimento}, {\bf 120}, N. 4,
353-366 (2005).

[17] A. Yu. Khrennikov,   {\it Phys. Lett. A} {\bf 316}, 279-296
(2003); {\it Annalen der Physik} {\bf 12},  575-585 (2003).

[18] A. Einstein, {\it The collected papers of Albert Einstein}
(Princeton Univ. Press, Princeton, 1993).

[19] A. Einstein and L. Infeld, {\it The evolution of Physics. From
early concepts to relativity and quanta} (Free Press, London, 1967).

[20] N. N. Bogolubov and N. N Bogolubov (son), {\it Introduction to
quantum statistical mechanics.} Nauka (Fizmatlit): Moscow (1984).

[21] A.Yu. Khrennikov, Infinite-Dimensional equation of Liuville.
{\it Mat. Sbornik}, {\bf 183}, 20-44 (1992).

[22] A. Yu. Khrennikov, The principle of correspondence in quantum
theories of field and relativistics bosonic string. {\it Mat.
Sbornic}, {\bf 180}, 763-786 (1989);  {\it Supernalysis.} Nauka,
Fizmatlit, Moscow, 1997 (in Russian). English translation: Kluwer,
Dordreht, 1999.

[23] J. S. Bell, {\it Speakable and unspeakable in quantum
mechanics.} Cambridge Univ. Press (1987).

[24] L. E. Ballentine, {\it Rev. Mod. Phys.}, {\bf 42}, 358--381
(1970).

[25] A. S. Holevo, {\it Statistical structure of quantum theory,}
Springer, Berlin-Heidelberg (2001).

\end{document}